\documentclass[fleqn,usenatbib]{mnras}

\usepackage{newtxtext,newtxmath}
\usepackage{multicol}
\usepackage[T1]{fontenc}
\usepackage{ae,aecompl}

\usepackage{colortbl}
\usepackage{hyperref}
\usepackage{lineno}
\usepackage[utf8]{inputenc}
\usepackage{dutchcal}
\usepackage{pgfplots}
\usepackage{subfigure}
\usepackage{siunitx}
\usepackage[percent]{overpic}
\usepackage{tikz}
\usepackage{natbib}
\usepackage[mmddyyyy]{datetime}
\usepackage{booktabs}
\usepackage[titletoc,title]{appendix}
\usepackage{subfigure}
\usepackage{graphicx}	
\usepackage{amsmath}	
\usepackage{amssymb}	
\usepackage{amsfonts}
\usepackage{amssymb}
\usepackage{etoolbox}
\makeatletter
\patchcmd\@combinedblfloats{\box\@outputbox}{\unvbox\@outputbox}{}{%
	\errmessage{\noexpand\@combinedblfloats could not be patched}%
}%
\makeatother
\modulolinenumbers[5]
\numberwithin{equation}{section}
\usetikzlibrary{calc}
\usetikzlibrary{plotmarks}
\usetikzlibrary{positioning}
\usetikzlibrary{fit}
\usetikzlibrary{decorations.pathmorphing}
\usetikzlibrary{backgrounds}
\pgfplotsset{compat = newest}
\pgfplotsset{ legend style={font=\tiny} }
\definecolor{bgreen}{rgb}{0.0,0.5,0.0}
\definecolor{bblue}{rgb}{0.0,0.0,0.9}
\definecolor{bgold}{rgb}{0.7,0.5,0.0}
\definecolor{bred}{rgb}{0.9,0.0,0.0}

\title[Truncated stellar kinetics]{A generalized Landau kinetic equation for \emph{weakly-coupled} probability distribution of $N$-stars in dense star cluster}
\author[Y. Ito]{
Yuta Ito$^{1,2,3}$\thanks{E-mail: yito@gradcenter.cuny.edu}
\\
$^{1}$Department of Physics, CUNY Graduate Center, 365 Fifth Avenue, New York, NY 10016, USA\\
$^{2}$Department of Engineering Science and Physics,
College of Staten Island, 2800 Victory Boulevard, Staten Island, NY 10314, USA\\
$^{3}$Department of Mathematics,
College of Staten Island, 2800 Victory Boulevard, Staten Island, NY 10314, USA
}
\date{Accepted XXX. Received YYY; in original form ZZZ}
        
\pubyear{2018}

\begin{document}
\label{firstpage}
\pagerange{\pageref{firstpage}--\pageref{lastpage}}
\maketitle

\begin{abstract}
The secular evolution of a collisional star cluster of $N$-'point' stars have been conventionally discussed based on cumulative two-body relaxation process. The relaxation process requires a cut-off on the range of two-body encounter between stars in physical space and the relaxation time is characterized by Coulomb logarithm $\ln[N]$; the conventional cut-off on the encounter distance in the literature gives "dominant" effect. In addition, incorrect cut-offs exposed a mathematical "infinite-density" problem in the late stage of core-collapse. 

The present paper shows these are merely the results due to incorrect cut-off process. If one correctly constrains the cut-off on interaction range between stars based on truncated BBGKY hierarchy, one must introduce a self-consistent 'truncated' Newtonian mean-field (m.f.) acceleration of star at position $\bmath{r}$ and time $t$ due to a phase-space distribution function $f\left(\bmath{r}', \bmath{p}',t\right)$ for stars	
\begin{align}
&\bmath{A}^{\triangle}(\bmath{r},t)=-Gm\left(1-\frac{1}{N}\right)\int_{\mid\bmath{r}-\bmath{r}' \mid > \triangle}\frac{\bmath{r}-\bmath{r}'}{\mid \bmath{r}-\bmath{r}' \mid^{3}} f\left(\bmath{r}',\bmath{p}',t\right)\text{d}^{3}{\bmath{r}'}\text{d}^{3}{\bmath{p}'},\nonumber
\end{align}
where $G$ is the gravitational constant and $m$ the mass of stars. The lower limit $\triangle$ of the distance between two stars is order of the Landau distance. 

The present paper shows the effect of total number on the structure of finite star cluster (in which stars undergo only two-body encounters) for the first time after establishing a mathematical formulation of a generalized Landau kinetic equation that includes the cut-off effect on both of collision term and m.f. potential by employing a BBGKY hierarchy for truncated DF stars. The cut-off effect increases the typical relaxation time by a few of percentage, which means the effect of cut-off itself is "non-dominant" on the relaxation time. On the other hand, the cut-off on m.f. acceleration is necessary to avoid "infinite-density" problem at the center of the system; the effect of total number $N$ on density profile and m.f. acceleration are shown by applying the truncated-DF BBGKY hierarchy to a toy model (a quasi-static modified Hubble density profile) for a core-halo structure of a star cluster at the late stage of evolution.  
\end{abstract}






\section{Introduction}\label{sec:intro}
A general point of view to understand statistical dynamics of dense star clusters is to introduce the effect of 'discreteness' of the clusters. The discreteness means the finiteness of total number $N$ of stars in a dense star cluster, say  $N\approx 10^{5}\sim 10^{7}$. In the present paper, the system of concern is collisional star clusters, e.g. globular clusters and collisional nuclear star clusters without super massive black holes. As a first approximation $(N\to\infty)$, the system can be assumed smooth and its evolution is dominated by a self-consistent mean field (m.f.) potential. The effect of m.f. potential is of significance on a few of dynamical-time scales and may freeze the system into a quasi-stationary state due to rapid fluctuations in m.f. field potential (i.e. violent relaxation). The evolutions of long-lived star clusters might have been driven by less probable relaxation process, two-body close encounters \citep[e.g.][]{Goodman_1983,Goodman_1984} ), and 'slow' many-body relaxations, statistical acceleration of stars and gravitational polarization \citep{Gilbert_1968}, in addition to the effect of m.f. potential\footnote{The present work focuses on systems modeled by kinetics of one-body distribution function of stars ('point particles' interacting via pair-wise Newtonian forces), neglecting the effect of triple encounters and some realistic effects (gas/dust/dark-mater dynamics, stellar evolution, inelastic direct collisions, formation of stars and binaries, stellar mass distribution, ...)}. 

The most fundamental relaxation process in the evolution of collisional star clusters is arguably the statistical acceleration that stands for a non-collective relaxation and mathematically modeled by generalized Landau kinetic equation \citep{Kandrup_1981,Chavanis_2013a}. The statistical acceleration originates from the deviation of the actual force on 'test' star due to ($N-1$)-'field' stars from the smooth force due to the m.f. potential \citep{Kandrup_1988}. The statistical acceleration may be considered in association with the effect of stochastic many-body encounters \citep{Kandrup_1981_a}. Conventionally, the effect of many-body encounters approximately gives place to that of cumulative \emph{two-body} encounters between stars \citep{Chandra_1942}. While the cumulative two-body encounters become more probable on larger-space scales due to the long-range nature of Newtonian pair-wise potential, the statistical acceleration becomes greater in magnitude on smaller scales. This implies the relaxation effects on intermediate-space scales are of significance in evolution of the system. As a matter of fact, the basic assumption made in use of stochastic kinetic equations\footnote{The stochastic kinetic equations here mean collision-Boltzmann \citep{Ipser_1983}, forward Komologouv-
Feller \citep{Kandrup_1980}, master\citep{Heggie_2003,Binney_2011,Merritt_2013}, Fokker-Planck\citep{Henon_1961} kinetic equations whose collision terms describe local two-body encounter in physical spaces with typically homogeneous background approximation.} is that 'test' star does not approach a 'field' star closer than the Landau distance and not go away far from the system size.  The cut-off on the range of effectve encounter-distance gives the follwoing estimation for the order of relaxation time\citep[e.g.][]{Ambartsumian_1938,Cohen_1950,Spitzer_1988}
\begin{align}
\frac{t_\text{r}}{t_\text{d}}\approx \frac{N}{\ln\left[p_\text{max}/p_\text{min}\right]}\sim \frac{N}{[\ln{ N}]} .\label{Eq.lnN_loc}
\end{align}
where $p_\text{max}$ is the maximum parameter, a typical size of clusters (tidal radius, King radius, Jean length ...), and $p_\text{min}$ is the minimum impact parameter, the Landau distance (typically independent of relative velocity between two stars).  The factor  $[\ln{ N}]$ stands for a parameter of relaxation time scale and corresponds with `Coulomb logarithm'. The logarithm has been employed as a measure of `finite-N' effect and many-body encounter \citep{Aarseth_1998} and the corresponding Fokker-Planck models have been a sucess in sense that it is a simple numerical method for stellar dynamics \citep{Heggie_2003,Binney_2011}
 
\subsection{A cut-off problem in use of kinetic theories}
Since the stochastic kinetic theoreis do not self-consistently include the effects of inhomgeneity (even m.f. potential) and collective effects of star clusters, for correct treatment of the matter, one must resort to the first principles; BBGKY hierarchy \citep{Gilbert_1968,Gilbert_1971,Chavanis_2013a} and Klimontovich-Dupree equation \citep{Chavanis_2012}. The formulation based on the first principles shows a Coulomb Logarithm in relaxation time for local encounters but in term of wavenumber  \citep[e.g.][]{Severne_1976, Kandrup_1981,Chavanis_2013a}
\begin{align}
\ln\left[\frac{k_\text{min}}{k_\text{max}}\right]\approx \ln[N].\label{Eq.lnN_non_loc}
\end{align}
where conventionally the following relations are assumed
\begin{subequations}
\begin{align}
&k_\text{min}\approx p_\text{max}\\
&k_\text{max}\approx p_\text{min}
\end{align}
\end{subequations}

Yet,  the fundamental assumption, equation \eqref{Eq.lnN_loc}, made for stochastic kinetic theory has not been `converted' into a self-consistent kinetic equation. The motivation for this work originates not only from the generalization work of the previous works but from some doubt for equations derived from stochastic kinetic theories and first principles. Use of stochastic kinetic equation allows one to employ typical m.f. potential that is smooth limitless in physical space; this is obviously inconsistent with the cut-off, equation \eqref{Eq.lnN_loc}. There are two kinds of test star exists in the system; a `uncorrelated' test star can approach a field star limitlessly forming a m.f. potential while `correlated' test star can not approach closer than the Landau distance to avoid close encounters. the outcome is the mathematical production of infinite density due to core collapse at the late stage of two-body relaxation evolution. On one hand, the equations derived based on the first principles also have a inconsistency. To find the relation \eqref{Eq.lnN_non_loc} the previous works assumes that test star can approach field star limitlessly while this is against assumptions of weak-coupling approximation and to be cut-off on scales of the Landau distance. 

The purpose of the present work is to derive based on a first principle a kinetic equation that correctly `cut-off' the encounter distance in physical space and the resolve the inconsistence of the existing kinetic theories. To do so the basic `target' of kinetic equation is the g-Landau kinetic equation. This is since the equation is known to correctly take into account the inhomogeneity effect and one does not have to assign cut-off on the maximum encounter-distance \citep{Kandrup_1988,Chavanis_2013a}. Yet, one needs to assign a lower cut-off for the encounter distance. One can resort to the use of BBGKY hierarchy truncated DF invented by \citep{Grad_1958} that can isolate a physical space where the short-range interaction between particles dominates from where weak-interaction occurs. The present work employes the truncated DF to cut-off the encounter-distance at Landau distance. This corresponds to a direct extension work of \citep{Takase_1950} where the Holtsmark distribution of force fields are employed and the strong-two body encounters and the formation of binaries are neglected by truncating the DF at order of Landau distance, termed as 'rough approximation'.

The present paper is organized as follows. In section \ref{sec:truncated_BBGKY} the truncated DF and the BBGKY hierarchy are explained. In sections \ref{sec:complete_WC} the g-Landau kinetic equation for the truncated DF of stars for a weakly-coupled star cluster is derived. In section \ref{sec:grainess_effect} the effect of cut-off on the m.f. potential and collision term, Coulomb logarithm, is discussed.  Section \ref{sec:conclusion} is Conclusion. 
\section{BBGKY hierarchy for truncated distribution function and non-ideal theory}\label{sec:truncated_BBGKY}
In section \ref{subsec:trunc_DF} truncated DF is  arranged for finite system and  in section \ref{sec:WC_star} 'weakly-coupled' DF is introduced to correctly includes the effect of cut-off on the interaction range of encounters. Section \ref{subsec:BBGKY} shows the BBGKY hierarchies for the DFs. 

\subsection{Truncated distribution function}\label{subsec:trunc_DF}
The truncated DF was originally introduced by \cite{Grad_1958} to derive the collisional Boltzmann equation for rarefied gases of particles interacting each other via short-range interaction of an effective potential distance $\triangle$. In the outside of sphere of radius $\triangle$ around test particle, one assumes no two-body interaction with a field particle occurs, or the pair-wise potential is much weaker than the inside of the sphere. The deficiency of the truncated DF, being not symmetric about permutation between the states of two stars, was improved in \citep{Cercignani_1972,Cercignani_1988} where the BBGKY hierarchies for the truncated DF of hard spheres and particles interacting via short-range pair potential were derived. The advantage of exploiting the truncated DF is three fold; (i) Among various derivations of the Boltzmann collision term, only the \citep{Grad_1958}'s method has a mathematically strict limit (Boltzmann-Grad limit); the ratio of particle size to the total particle numbers as proved in \citep{Lanford_1981} and can avoid the mathematical divergence problem at the $N$-body Liouville-equation level (ii) The \citep{Grad_1958}'s method allows one to derive a kinetic equation in spherical coordinates; one can discuss the effect of two-body encounters and the statistical acceleration in the same coordinates\footnote{The wave kinetic theories are in general discussed in spherical coordinates in terms of relative displacement between two stars, while collision ones typically assumes cylindrical coordinates (Appendix \ref{Appendix:Bogorigouv}).} (iii) Statistical dynamics of two-body encounter can be separated at $r_{ij}=\triangle$ from the deterministic Newtonian mechanics inside the Landau sphere \citep[e.g.][]{Cercignani_2008}.

An $s$-tuple truncated DF of stars may be defined as\footnote{\cite{Cercignani_1972,Cercignani_1988} used the $s$-body (symmetric) joint-probability DF and the Boltzmann-Grad limit ($N\triangle^{2}\rightarrow\mathcal{O}(1)$ as $N\to\infty$), meaning the small number $s$ in the factor $\frac{N!}{(N-s)!}$ is less important, while stellar dynamics necessitate the small $s$ to discuss the granularity. Accordingly, the formulas shown in the present work are slightly different from the Cercignanni's work due to the definition for DF.}
\begin{equation}
f^{\triangle}_{s}(1, \cdots, s, t)=\frac{N!}{(N-s)!}\int_{\Omega_{s+1,N}}F_N(1,\cdots,N, t)\text{d}_{s+1}\cdots\text{d}_N,\label{Eq.SbodytruncDF}
\end{equation} 
where  $1\lid s \lid N-1$. The effective interaction range $\triangle$ throughout the present paper is considered as the Landau radius $r_\text{o}$ defined by
\begin{align}
r_\text{o}\equiv\frac{2}{1+\sqrt{2}}\frac{Gm}{<\varv>^{2}}.\label{Eq.thermo_approx_r}
\end{align}
and $r_\text{o}$ means the closest separation of stars in two-body encounter under dispersion approximation (See Appendix \ref{subsec:basic_scaling} for the detail definition.)
 
 In equation \eqref{Eq.SbodytruncDF} the $F_{N}$ is $N$-body joint-probability DF  i.e. the phase-space probability density of finding stars $1, 2, \cdots, N$ at phase-space points $(\bmath{r}_{1},\bmath{p}_{1})$, $(\bmath{r}_{2},\bmath{p}_{2})$, $\cdots$ and $(\bmath{r}_{N},\bmath{p}_{N})$ respectively at time $t$. The arguments $\{1,\cdots,N\}$ of the $N$-body DF $F_{N}$ are the position coordinates and momenta $\{\bmath{r}_{1},\bmath{p}_{1},\cdots,\bmath{r}_{N},\bmath{p}_{N}\}$ of stars in the system.  The $N$-body DF is normalized as
\begin{equation}
\int F_N(1,\cdots,N, t)\text{d}_{1}\cdots\text{d}_{N}=1,\label{Eq.normalizedDF}
\end{equation}
where an abbreviated notation is employed for the phase-space volume elements, $\text{d}_{1}\cdots\text{d}_{N}(=\text{d}\bmath{r}_{1}\text{d}\bmath{p}_{1}\cdots\text{d}\bmath{r}_{N}\text{d}\bmath{p}_{N})$. In addition, the function $F_{N}$ is assumed symmetric about a permutation between any two phase-space states of stars \citep{Balescu_1997,Liboff_2003}.

Equation \eqref{Eq.SbodytruncDF} is in essence the same as the definition for the truncated DF used in \citep{Cercignani_1972} though, it has a reduced form since the following $s$-body DFs is symmetric in permutation between two phase-space states of stars; 
\begin{equation}
f_{s}(1\cdots s, t)=\frac{N!}{(N-s)!}F_{s}(1\cdots s, t).\label{Eq.StupleDF}
\end{equation} 
The $s$-tuple DF describes the probable number (phase-space) density of finding stars $1,2,\cdots,s$ at phase-space points $1,2,\cdots,s$ respectively. The domain of integration in equation \eqref{Eq.SbodytruncDF} must be taken over the limited phase-space volumes $\Omega_{s+1,N}$ defined by
\begin{equation}
\Omega_{s+1,N}=\left(\{\bmath{r}_{s+1},\bmath{p}_{s+1}\cdots\bmath{r}_{N},\bmath{p}_{N}\}\Bigg\vert \prod_{i=s+1}^{N}\prod_{j=1}^{i-1}\left\{{\mid \bmath{r}_{i}-\bmath{r}_{j} \mid > \triangle}\right\}\right).\label{Int_Omega}
\end{equation}
For example,
\begin{subequations}
	\begin{align}
	&\Omega_{2,2}=\left(\{\bmath{r}_{2},\bmath{p}_{2}\} \Big\vert\{\mid \bmath{r}_{1}-\bmath{r}_{2} \mid > \triangle\}\right),\\
	&\Omega_{3,3}=\left(\{\bmath{r}_{3},\bmath{p}_{3}\}\Big\vert\{\mid \bmath{r}_{1}-\bmath{r}_{3} \mid > \triangle\}\times\{\mid \bmath{r}_{2}-\bmath{r}_{3} \mid > \triangle\}\right),
	\end{align}
\end{subequations}
and refer to Appendix \ref{Appendix_BBGKY} for more detail discussion. 
\begin{table}
	\caption{A schematic description of the truncated DF. Kinetic description for stars $1,2,\cdots,k,\cdots, N$ follows the wave kinetic description unless one of stars enters the Landau sphere of another star while the collision kinetic description must be employed if any two stars approaches closer than the Landau radius.}
	\begin{tikzpicture}
	\shade[ball color = gray!40, opacity = 0.4] (0,0) circle (1.41421cm);
	\draw (0,0) circle (1.41421cm);
	\draw (-1.41421,0) arc (180:360:1.41421 and 0.6);
	\draw[dashed] (1.41421,0) arc (0:180:1.41421 and 0.6);
	\draw[dashed] (0,0 ) -- node[above]{$\triangle$} (1.41421,0);
	\shadedraw[inner color=orange, outer color=yellow, draw=black] (0,0) circle (0.1cm);
	\node [below] at (0,0.0) {star 1};	
	
	\shade[ball color = gray!40, opacity = 0.4] (4.5,-3) circle (1.41421cm);
	\draw (4.5,-3) circle (1.41421cm);
	\draw (3.11421,-3) arc (180:360:1.41421 and 0.6);
	\draw[dashed] (5.91421,-3) arc (0:180:1.41421 and 0.6);
	\draw[dashed] (4.5,-3 ) -- node[above]{$\triangle$} (5.91421,-3);
	\shadedraw[inner color=orange, outer color=yellow, draw=black] (4.5,-3.0) circle (0.1cm);
	\node [below] at (4.5,-3.0) {star 2};
	
	\shade[ball color = gray!40, opacity = 0.4] (4,1) circle (1.41421cm);
	\draw (4,1) circle (1.41421cm);
	\draw (2.61421,1) arc (180:360:1.41421 and 0.6);
	\draw[dashed] (5.41421,1) arc (0:180:1.41421 and 0.6);
	\draw[dashed] (4,1) -- node[above]{$\triangle$} (5.41421,1);
	\shadedraw[inner color=orange, outer color=yellow, draw=black] (4,1) circle (0.1cm);
	\node [below] at (4.0,1) {star $k$};
	
	\end{tikzpicture}
\end{table}
The truncated $s$-tuple DF $f^{\triangle}_{s}(1, \cdots, s, t)$ is assumed symmetric about a permutation between two states. The truncated single- and double- DFs explicitly read
\begin{subequations}
	\begin{align}
	&f^{\triangle}_{1}(1,t)\nonumber\\
	&=f_{1}(1,t)-\frac{1}{2}\int_{r_{23} < \triangle}f_{3}(1,2,3,t)\text{d}_{2}\text{d}_3-\int_{r_{12} < \triangle}f_{2}(1,2,t)\text{d}_{2}\nonumber\\
	&\quad+\left[i\int_{\substack{r_{12} < \triangle\\\times r_{13}< \triangle}}+\iint_{\substack{r_{12}  < \triangle\\\times r_{23}< \triangle}}-\iint_{\substack{r_{12}  < \triangle\\ \times r_{13}  < \triangle\\\times r_{23}< \triangle}}\right]f_{3}(1,2,3,t)\text{d}_{2}\text{d}_{3}\nonumber\\
	&\qquad-...\label{Eq.single_truncated_DF}\\
	&f^{\triangle}_{2}(1,2,t)\nonumber\\
	&=f_{2}(1,2,t)-\frac{1}{2}\int_{r_{34} < \triangle}f_{4}(1,2,3,4,t)\text{d}_{3}\text{d}_4\nonumber\\
	&\quad-\left[\int_{\mid r_{13} \mid < \triangle}+\int_{\mid r_{23} \mid < \triangle}-\int_{\substack{\mid r_{13} \mid < \triangle\\\times\mid r_{23} \mid < \triangle}}\right]f_{3}(1,2,3,t)\text{d}_{3}\nonumber\\
	&\quad+\iint\text{d}_{3}\text{d}_{4} f_{4}(1,\cdots, 4,t)\nonumber\\
	&\qquad\times\{\Theta(\triangle-r_{13})[\Theta(\triangle-r_{14})+\Theta(\triangle-r_{23})+\Theta(\triangle-r_{34})]\nonumber\\
	&\qquad\quad+\Theta(\triangle-r_{23})[\Theta(\triangle-r_{24})+\Theta(\triangle-r_{34})]\nonumber\\
	&\qquad\quad-2\Theta(\triangle-r_{13})\Theta(\triangle-r_{23})\nonumber\\
	&\qquad\qquad\times[\Theta(\triangle-r_{14})+\Theta(\triangle-r_{23})+\Theta(\triangle-r_{34})]\}\nonumber\\
	&\qquad-...\label{Eq.double_truncated_DF}
	\end{align}
\end{subequations}
where $\Theta(\cdot)$ describes a Heaviside step function. Hence the truncated single (double) DF describes the probability $not$ to find star 1 (star 1 or 2) around star 2 (star 3) within the region inside a sphere of radius $\triangle$ (the Landau sphere) at time $t$. Despite of the mathematically strict definition for the truncated DFs, it does not have a straightforward physical meaning; one may resort to a simplification of the truncated DF\footnote{In \citep{Grad_1958,Cercignani_1972}, the interaction range essentially goes to zero due to the Boltzmann-Grad limit $\triangle\sim\frac{1}{\sqrt{N}} \to 0$ and the truncated DF is considered as a standard DF.}. Due to the shortness of the interaction range of  $\triangle$ between two stars
\begin{align}
\triangle\equiv r_\text{o}\sim\mathcal{O}\left(\frac{1}{N}\right),\label{Eq.Landau_distance}
\end{align}
the truncated single- and double- DFs can be approximated to
\begin{subequations}
	\begin{align}
	&f^{\triangle}_{1}(1,t)=f_{1}(1,t)-\frac{1}{2}\iint_{r_{23} < \triangle}f_{3}(1,2,3,t)\text{d}_{2}\text{d}_{3}+\mathcal{O}(1/N),\label{Eq.trunc_DF1_N}\\
	&f^{\triangle}_{2}(1,2,t)=f_{2}(1,2,t)-\frac{1}{2}\iint_{r_{34} < \triangle}f_{4}(1,2,3,4,t)\text{d}_{3}\text{d}_{4}+\mathcal{O}(1).\label{Eq.trunc_DF2_N}
	\end{align}\label{Eq.approx_truncated_DFs}
\end{subequations}
The second terms on the R.H.S of equations \eqref{Eq.trunc_DF1_N} and \eqref{Eq.trunc_DF2_N} show the effect of discreteness on the DFs. One should be aware of the effect of discreteness on the truncated DF being associated with the randomness (fluctuation in the m.f. potential \citep{Chandra_1943a,Takase_1950}) rather than that one generally discusses\footnote{It is obvious in stellar dynamics that a strict definition for typical DF itself is difficult to achieve due to the 'discreteness' or granularity of the system in phase space $(\bmath{r},\bmath{p})$. The 'discreteness' stands for 'sparse' physical infinitesimal elements of phase space \citep[][pg. 9]{Spitzer_1988}; what one can do is to take the DF in terms of integrals of motion and orbit-averaging it.}. The obvious complication of the DFs, equations \eqref{Eq.trunc_DF1_N} and \eqref{Eq.trunc_DF2_N}, may be comforted by excluding the possibility of triple encounter. In the Landau sphere of radius $r_{23}=\triangle$ or $r_{34}=\triangle$,  any stars other than the stars of concern (stars 2 and 3 or stars 3 and 4 respectively) can not exist in the Landau sphere under the two-body encounter approximation. Hence, equation \eqref{Eq.approx_truncated_DFs} can be reduced to
\begin{subequations}
	\begin{align}
	&f^{\triangle}_{1}(1,t)=f_{1}(1,t)\left(1-\frac{1}{2}\iint_{r_{23} < \triangle}f_{2}(2,3,t)\text{d}_{2}\text{d}_{3}\right),\label{Eq.trunc_DF1_N_2}\\
	&f^{\triangle}_{2}(1,2,t)=f_{2}(1,2,t)\left(1-\frac{1}{2}\iint_{r_{34} < \triangle}f_{2}(3,4t)\text{d}_{3}\text{d}_{4}\right).\label{Eq.trunc_DF2_N_2}
	\end{align}\label{Eq.approx_truncated_DFs_2}
\end{subequations}
The fundamental idea of truncated DF is that the truncation of phase-space volume makes the system 'open' on small scales. This may be clearly understood if one takes the integral $\int\cdot\text{d}_{1}$ over equation \eqref{Eq.trunc_DF1_N_2} and $\iint\cdot\text{d}_{1}\text{d}_{2}$ over equation \eqref{Eq.trunc_DF2_N_2};
\begin{subequations}
	\begin{align}
	&\int f^{\triangle}_{1}(1,t)\text{d}_{1}=N\left(1-\frac{1}{2}\iint_{r_{23} < \triangle}f_{2}(2,3,t)\text{d}_{2}\text{d}_{3}\right),\label{Eq.trunc_N1}\\
	&\iint f^{\triangle}_{2}(1,2,t)\text{d}_{1}\text{d}_{2}\nonumber\\
	&\qquad=N(N-1)\left(1-\frac{1}{2}\iint_{r_{34} < \triangle}f_{2}(3,4t)\text{d}_{3}\text{d}_{4}\right).\label{Eq.trunc_N2}
	\end{align}\label{Eq.approx_truncated_N}
\end{subequations}
The total number $N$ of stars described by the truncated DFs does not conserve since the DFs 'overlook' counting the probable number of stars in the Landau spheres(, which is useful only for binary formation and disruption/coalescence.). This obvious complication may be avoided by assuming two different assumptions. First, one may assume no star can approach another star than the Landau radius. Such stars will be termed \emph{weakly-coupled} (WC) stars in the present paper. The WC stars are mathematically defined in section \ref{sec:WC_star} and applied to a star cluster in section \ref{sec:complete_WC}. Second, one may also apply the 'test-particle' method \footnote{The 'test-particle' method means that only test star (star 1) can approach one of field stars closer than the Landau radius but none of the other field stars can, meaning one does not find any stars in the Landau sphere of radius $r_{23}=\triangle$ or $r_{34}=\triangle$. (It is to be noted whether star 1 is in the Landau sphere of star 2 or not is not a crucial discussion since it comes into a play at order of $1/N^{2}$ as seen in the third term on the R.H.S of equation \eqref{Eq.single_truncated_DF}.).} of \citep{Kaufman_1960,Kandrup_1981} to be explained in later paper. The both of assumptions (WC-stars approximation or the 'test-particle' method) can avoid the non-conservation of total number of stars;
\begin{subequations}
	\begin{align}
	&f^{\triangle}_{1}(1,t)=f_{1}(1,t)+\mathcal{O}(1/N^{2}),\label{Eq.trunc_DF1_test}\\
	&f^{\triangle}_{2}(1,2,t)=f_{2}(1,2,t)+\mathcal{O}(1/N).\label{Eq.trunc_DF2_test}
	\end{align}
\end{subequations}
Hence, the truncated $s$-tuple DFs of stars may be treated as the standard DFs, equations \eqref{Eq.singleDF} and \eqref{Eq.doubleDF}.

The total energy of $N$ stars of equal masses $m$ in a star cluster has the following forms in terms of the truncated DFs
\begin{align}
&E(t)^{\triangle}=\int \frac{\bmath{p}_{1}^{2}}{2m}f^{\triangle}(1,t)\text{d}_{1}+U^{\triangle}(t),\label{Eq.E_trun}
\end{align}
where
\begin{align}
U^{\triangle}(t)=\frac{m}{2}\int_{r_{12} > \triangle}\phi(r_{12})f^{\triangle}(1,2,t)\text{d}_{1}\text{d}_{2},\label{Eq.U_trun}
\end{align}
where $\phi(r_{ij})$ is the Newtonian gravitational potential due to star $j$ that star $i$ feels
\begin{equation}
\phi(r_{ij})=-\frac{Gm}{r_{ij}} \qquad(1\lid i, j \lid N \quad \text{with} \quad i\neq j ),\label{Eq.phi}
\end{equation}
where $G$ is the gravitational constant and $r_{ij}=|\bmath{r}_{i}-\bmath{r}_{j}|$ is the distance between stars $i$ and $j$.  In the same way as the non-conservation of total number of stars, the truncated DFs do not conserve the total energy. If one does not resort to any approximation, equation \eqref{Eq.E_grain} states even the total energy of a finite star cluster must be conserved only up to order of $\mathcal{O}(1)$. Hence, employing the WC-star approximation or 'test-particle' method, one obtains the total energy of stars outside the Landau spheres
\begin{align}
&E(t)^{\triangle}=\int \frac{\bmath{p}_{1}^{2}}{2m}f(1,t)\text{d}_{1}+U_{\text{m.f.}}^{\triangle}(t)+U_{\text{cor}}^{\triangle}(t),\label{Eq.E_grain}
\end{align}
where
\begin{subequations}
	\begin{align}
	&U_{\text{m.f.}}^{\triangle}(t)=\frac{m}{2}\int\Phi^{\triangle}(\bmath{r}_{1},t)f(1,t)\text{d}_{1},\label{Eq.U_id_grain}\\
	&U_{\text{cor}}^{\triangle}(t)=\frac{m}{2}\int_{r_{12} > \triangle}\phi(r_{12})g(1,2,t)\text{d}_{1}\text{d}_{2},\label{Eq.U_cor_grain}
	\end{align}
\end{subequations}
and the self-consistent \emph{truncated} m.f. potential is defined as 
\begin{align}
&\Phi^{\triangle}(\bmath{r}_{1},t)=\left(1-\frac{1}{N}\right)\int_{r_{12} > \triangle} \phi(r_{12})f(2,t)\text{d}_{2}.\label{Eq.Phi_grain}
\end{align}
The corresponding \emph{truncated} m.f. acceleration reads
\begin{align}
&\bmath{A}^{\triangle}(\bmath{r}_{1},t)=-\left(1-\frac{1}{N}\right)\int_{r_{12} > \triangle} \nabla_{1} \phi(r_{12})f(2,t)\text{d}_{2}.\label{Eq.A_grain}
\end{align}
One must recall that the DFs, equations \eqref{Eq.trunc_DF1_test} and \eqref{Eq.trunc_DF2_test}, inside the Landau sphere do not have a statistically strict meaning. The truncated m.f. potential, equation \eqref{Eq.Phi_grain}, and acceleration, equation \eqref{Eq.A_grain}, seem an artificial concept though, it gives a clear physical meaning. The truncated DF assigns a \emph{geometrical} constraint on a standard double DF (both of the product of uncorrelated DFs and correlation function) that the dynamics of stars (e.g. Newtonian two-body interaction, formation of binaries, coalescence and disruption) inside the Landau sphere does not 'coincide' with the statistical quantity at the same distance to describe the system, which corresponds with the 'rough approximation \citep{Takase_1950}' of randomness in Holtsmark DF. Hence, fluctuations in m.f. acceleration can be excited only outside the sphere. The truncated m.f. acceleration, equation \eqref{Eq.A_grain}, also stands for a case in which the m.f. acceleration of a star due to stars traveling in a Landau sphere does not contribute to the stellar dynamics. (Hence, the polarization across the surface of the Landau sphere must be ignored.). In section \ref{subsec:grainess_effect_Poisson}, the Poisson equation for the truncated DF of stars will be explained.

\subsection{'Weakly-coupled' Distribution Function}\label{sec:WC_star}
To avoid the non-conservation of total- number and energy of stars described by the truncated DF, in the present section, the hard-sphere DF \citep{Cercignani_1972} will be extended to the weakly-coupled DF of stars. \cite{Cercignani_1972} extended the Grad' truncated DF into the hard-sphere DF to derive the collisional Boltzmann equation for rarefied gases of hard-sphere particles. The hard-sphere model does not allow any particles of radii $\triangle$ exist inside the other particles of radii $\triangle$ in a rarefied gas; it is defined as
\begin{align}
&f_{N}^{\blacktriangle}(1,\cdots, N,t)=
\begin{cases}
f_{N}(1,\cdots,N,t),     & \text{if}\quad r_{ij}\geq \triangle \text{with}\quad i\neq j \\
0,              & \text{otherwise} \label{Eq.truncDf_HS}
\end{cases}
\end{align}
Following the definition of single- and double- truncated DFs, equations \eqref{Eq.single_truncated_DF} and \eqref{Eq.double_truncated_DF}, the first two $s$-tuple hard-sphere DFs explicitly read
\begin{subequations}
	\begin{align}
	&f_{1}^{\blacktriangle}(1,t)=f_{1}(1,t),\\
	&f_{2}^{\blacktriangle}(1,2,t)=
	\begin{cases}
	f_{2}(1,2,t),     & \text{if}\quad r_{12}\geq \triangle\\
	0,              & \text{otherwise} \label{Eq.truncDf_HS_12}
	\end{cases}
	\end{align}
\end{subequations}
In equation \eqref{Eq.truncDf_HS_12}, the hard-sphere double DF is smooth and continuous, well-defined as limit of $r_{12}\rightarrow\triangle^{+}$, while it can be discontinuous  as limit of $r_{12}\rightarrow\triangle^{-}$. Hence, the value of the double DF at the radius $r_{12}=\triangle$ is defined as the limit value 
\begin{align}
\left[f_{2}(1,2,t)\right]_{r_{12}=\triangle} = \lim_{r_{12}\to \triangle+}f_{2}^{\blacktriangle}(1,2,t). \label{Eq.truncDf_HS_lim}
\end{align}
On a star cluster if one assumes a strong constraint that any star can not approach any other stars closer than the Landau distance, equation \eqref{Eq.Landau_distance} (while the maximum separation between stars is bounded by the system size), the weak-coupling approximation may be actually embodied:
\begin{align}
&r_\text{o}\qquad  < \quad r_{12}\quad \leq\qquad  R, \label{Eq.weak_coupling}\\
&\mathcal{O}\left(\frac{1}{N}\right) \qquad\qquad\qquad\quad \mathcal{O}(1)\nonumber
\end{align}
This ideal mathematical condition is interpreted as an extreme case of the hard-sphere DF, equation\eqref{Eq.truncDf_HS}, with the limit value of zero for $f(1,2,t)$ at $r_{12}=\triangle$;
\begin{align}
\left[f_{2}(1,2,t)\right]_{r_{12}=\triangle} = 0.\label{Eq.WC_r12}
\end{align}
and the corresponding definition for the heviside funciton is uniquely determined\footnote{Yet, the present work relies on the formulation based on Heaviside function and derivatives, at least to hold the Lebnitz rule, one needs to employ the following defintion
	\begin{align}
	\Theta(r_{ij}-\triangle)\equiv\begin{cases}1\qquad (r_{ij}>\triangle)\\
	\frac{1}{2}\qquad (r_{ij}=\triangle)\\
	0 \qquad (r_{ij}<\triangle)
	\end{cases}
	\end{align}
	This formulation is of significance \emph{only} to hold the surface integral terms one may assume that equation \eqref{Eq.weak_coupling} is true. Accoringly, use of equation \eqref{Eq.Def_step_func} prohibits one to take a derivative of any product of identical step functions with respect to $\bmath{r}_{1}$, $\bmath{r}_{2}$ and $\bmath{r}_{12}$; this may be possible at BBGKY-hierarchy level since the $s$-tuple DFs (inculding correlation functions) are linealy independent.} as follows
\begin{align}
\Theta(r_{ij}-\triangle)\equiv\begin{cases}1\qquad (r_{ij}>\triangle)\\
0 \qquad (r_{ij}\leq\triangle) \label{Eq.Def_step_func}
\end{cases}
\end{align}
The hard-sphere DF with the condition, equation \eqref{Eq.WC_r12}, is termed a weakly-coupled DF in the present work to isolate itself from hard-sphere DF. The weakly-coupled DF in essence corresponds with the 'Rough approximation \citep{Takase_1950}' of the random factor for the Holtsmanrk distribution of Newtonian force strength, meaning the relative velocity dependence between test- and a field- star will be neglected when the test star entering the Landau sphere in the present work for simplicity.

\subsection{Truncation Condition for Weakly-coupled DF}
The definition for weakly-coupled DF gives the thresh point $(r_{12}=\triangle)$ a physical causality in space, i.e. a direct collision between two spheres occurs only from the outside of each sphere. One must be careful to deal with the explicit form of double or higher order of $s$-tuple hard-sphere DF. The double DF may be explicitly defined as
\begin{subequations}
	\begin{align}
	f_{2}^{\blacktriangle}(1,2,t)&=
	\begin{cases}
	\left(1-\frac{1}{N}\right)f(1,t)f(2,t)+g(1,2,t),     & \text{if}\quad r_{12}\geq \triangle\\
	0,              & \text{otherwise} \label{Eq.truncDf_HS_expl}
	\end{cases}\\
	&\equiv \left(1-\frac{1}{N}\right)\left[f(1,t)f(2,t)\right]_{r_{12}\geq \triangle}+g(1,2,t)_{r_{12}\geq \triangle},\label{Eq.hard_f(1,2,t)}
	\end{align}
\end{subequations}
where the DFs $f(1,t)$ and $f(2,t)$ are not exactly statistically uncorrelated since the geometrical condition assigned on the interaction range, $r_{12}>\triangle$, must be considered; only the DFs $f^{\blacktriangle}(1,t)$ and $f^{\blacktriangle}(2,t)$ are statistically independent each other. Hence,
\begin{align}
\left[f(1,t)f(2,t)\right]_{r_{12}\geq \triangle}\neq f^{\blacktriangle}(1,t)f^{\blacktriangle}(2,t). \label{Eq.neq_DFs}
\end{align}
Also, the hard-sphere DF is different from the DF, equation \eqref{Eq.SbodytruncDF}, in sense that the phase-space domain of truncated DF is limited always through that of integration, while hard-sphere does not have domain itself in the Landau sphere. To specify the explicit form of DFs, one may employ the following form
\begin{align}
&f^{\blacktriangle}(1,2,t)\equiv\Theta(r_{12}-\triangle)f(1,2,t)\label{Eq.def_WC_DF2}\\
&f^{\blacktriangle}(1,2,3,t)\equiv\Theta(r_{12}-\triangle)\Theta(r_{13}-\triangle)\Theta(r_{23}-\triangle)f(1,2,3,t)\label{Eq.def_WC_DF2}
\end{align}
For self-consistent relation, the total number is
\begin{align}
\int_{r_{12}>\triangle}f^{\blacktriangle}(1,2,t)\text{d}2=\left(N-1\right)f(1,t)\left( 1-\frac{1}{N} \int_{r_{12}<\triangle} f(2,t)\text{d}2 \right)\label{Eq.int_trinc_twoDF}
\end{align}
To hold the consistent relation between DF $f(1,t)$ and higher orders of DF, one may consider two cases (i) approximated form of DF and (ii) exact form of weakly-coupled DF. The two cases are discussed in sections 

\subsubsection{Approximated form of weakly-coupled DF}
To employ standard $f(1,t)$, one may approximate the second term on the R.H.S of equation \eqref{Eq.int_trinc_twoDF} to
\begin{align}
&\int_{r_{12}>\triangle}f^{\blacktriangle}(1,2,t)\text{d}2=\left(N-1\right)f(1,t)+\mathcal{O}(\frac{1}{N})\label{Eq.int_trinc_twoDF_approx}\\
&\int\Theta(r_{13}-\triangle)\Theta(r_{23}-\triangle)f^{\blacktriangle}(1,2,3,t)\text{d}3=\left(N-2\right)f(1,2,t)+\mathcal{O}(1)\label{Eq.int_trinc_threeDF_approx}
\end{align}
where the second terms on the R.H.S of equations are $N^{3}$ times weaker than the first term in order of magnitude. This implies one may employ the definition, equation \eqref{Eq.def_WC_DF}, till the density of system reaches $N^{2}\bar{n}$ where $\bar{n}$ is the (initial) mean density of the system. Also, one may employ standard definition for total number of stars.
\begin{align}
N=\int f(1,t)\text{d}1
\end{align}
\subsubsection{Exact form of weakly-coupled DF}
The straightforward but hard-to-accept way to employ weakly-coupled DF is to employ the following definition for total number of stars
\begin{align}
\int_{r_{11'}>\triangle}f(1',t)\text{d}1'=N\equiv N^{*}(\bmath{r}_{1},t)
\end{align}
Accordingly, a correct definition for DFs for stars $2\cdots N$
\begin{align}
&\int\Theta(r_{13}-\triangle)\Theta(r_{12}-\triangle)f(3,t)\text{d}3=N\\
&\qquad\qquad \vdots\\
&\int\Theta(r_{13}-\triangle)\cdots\Theta(r_{1N}-\triangle)f(N,t)\text{d}N=N
\end{align}

This definition necessitates ones to consider change of the total number $ N*(\bmath{r}_{1},t)$ with time $t$
\begin{align}
\frac{\text{d}N^{*}(\bmath{r}_{1},t)}{\text{d}t}&=\int_{r_{11'}>\triangle}\frac{\partial f(1',t)}{\partial t}\text{d}1'\\
&=-\frac{\bmath{\varv}_{1}\cdot\hat{r}_{1}\triangle}{3}a_{1}(r_{1},\triangle)
\end{align}
where 
\begin{align}
a_{1}(r_{1},\triangle)\hat{r}_{1}=\frac{3}{4\pi}\int n(\bmath{r}_{1}-\triangle\hat{r'},t)\hat{r'}\text{d}\Omega'
\end{align}
The obvious condition to hold the total number is the factor $a_{1}(r_{1},\triangle)$ vanishes. This is the case when the system does not have a peculiar structure i.e. no change in density with spatial translation $n(\bmath{r}_{1}-\triangle)\approx n(\bmath{r}_{1})$ on scale of the Landau radius 
\begin{align}
a_{1}(r_{1},\triangle)\hat{r}_{1}\approx\frac{3}{4\pi}n(\bmath{r}_{1},t)\int \hat{r'}\text{d}\Omega'=0\label{Eq.conserv_condit}
\end{align}
Also one may consider the 'conservation' of total number with space
\begin{align}
\nabla_{1}N^{*}(\bmath{r}_{1},t)=\frac{4\pi\triangle}{3}a_{1}(r_{1},\triangle)\hat{r}_{1}
\end{align}
This can vanish due to the condition \eqref{Eq.conserv_condit}.

\subsection{rough approximation of randomness factor for stars entering the Landau sphere  }\label{subsec:rough_approx}
In the present work following the rough approximation in Chandra 1941 Takase 1950   where the interaction range two-body encounter was limited and neglects the relative velocity of the stars entering the Landau sphere. If one introduces the randomness factor, following Takase1950 for a homogeneous static background, by separating the relative speed into speeds associated with deterministic two-body encounter and randomness fluctuation
\begin{align}
&\chi(r)=\int^{\infty}_{\varv_{12}(-\infty)}\chi'(\bmath{r},\bmath{\varv})f(\bmath{\varv})\text{d}\varv\\
&\chi'(\bmath{r},\bmath{\varv})=\frac{1}{4\pi}\int^{2\pi}_{0}\int^{\pi-\theta_{1}}_{0}\sin\theta\text{d}\theta\text{d}\psi\\
&\sin\theta_{1}=\frac{r_\text{o}}{r}
\end{align}
where the angle $\theta_{1}$ is azimuthal angle forming a cone atop which test star enter the deterministic region and at bottom which a circle of radius $r_\text{o}$ with center of a filed star. If one considers that the DF of stars with relative speeds has a Maxwellian
\begin{align}
f(\varv)=\frac{<\varv>^{3}}{\pi^{3/2}}\text{e}^{-\varv^{2}/<\varv>^{2}}
\end{align}
then one can  approximate the function $\chi(r)$  to 
\begin{align}
\chi_\text{app}(r)\approx\frac{1}{1+(a_\text{B.G.}/r)^{2}}\qquad (r>>r_\text{o})
\end{align}
where $a_\text{B.G.}$ is `encounter radius Ogornogouv' above which one may consider the effect of m.f. acceleration is dominant, meaning one can assume a periodicity of orbits of stars and fluctuations in m.f. potential. What one must  discuss here is ``what portion of stars may be less counted if one neglects the stars feeling deterministic interaction''. This can estimated by
\begin{align}
\left(1-\frac{N}{\bar{n}}\int^{R}_{0}\chi_\text{app}\text{d}\bmath{r}\right)&=2\left(\frac{a_\text{B.G.}}{R}\right)^{3}-3\left(\frac{a_\text{B.G.}}{R}\right)^{6}\ln\left[\frac{R}{a}\right]\\
&\approx \mathcal{O}\left(\frac{1}{N^{3/2}}\right)
\end{align}
Since the `small' number of concern in the present work is $\sim 1$ to pick up the effect of `discreteness' at kinetic-equation level compared to $f(1,t)\sim N$, the stars interacting without randomness may be put aside from the main discussion of concern. Hence use of rough approximation for randomness factor may be granted as done in chandra Takase  and 
\begin{align}
\chi_\text{rou}=\Theta(r-r_\text{o})
\end{align}
which corresponding to the truncation of DF in the present work.

\subsection{BBGKY hierarchies for standard, truncated and hard-sphere DFs}\label{subsec:BBGKY}
In a very similar way to the derivation of standard BBGKY hierarchy, the BBGKY hierarchy for the truncated $s$-body function can be found (refer to Appendix \ref{Appendix_BBGHY}, or see \cite{Cercignani_1972,Cercignani_1988}) as 
\begin{align}
&\partial_{t}f^{\triangle}_{s}+\sum_{i=1}^{s}\left[\bmath{\varv}_{i}\cdot\nabla_{i}+\sum_{\quad{j=1(\neq i)}}^{s}\bmath{a}_{ij}\cdot\bmath{\partial}_{i}\right]f^{\triangle}_{s}\nonumber\\
&\qquad +\sum_{i=1}^{s}\left[\bmath{\partial}_{i}\cdot\int_{\Omega_{s+1,s+1}}f^{\triangle}_{s+1}\bmath{a}_{i,s+1}\text{d}_{s+1}\right]\nonumber\\
&=\sum_{i=1}^{s}\left[\int\text{d}^{3}\varv_{s+1}\oiint f^{\triangle}_{s+1} \bmath{\varv}_{i,s+1}\cdot\text{d}\bmath{\sigma}_{i,s+1}\right]\nonumber\\
&\qquad+\frac{1}{2}\int\text{d}^{3}\varv_{s+2}\int\text{d}_{s+1}\oiint f^{\triangle}_{s+2}  \bmath{\varv}_{s+1,s+2}\cdot\text{d}\bmath{\sigma}_{s+1,s+2},\label{Eq.BBGKY_truncated}
\end{align}
where the relative velocity $\bmath{\varv}_{ij}$ and the acceleration $\bmath{a}_{i}$ of star $i$ due to the `potential' force from the rest of stars are defined as
\begin{subequations}
\begin{align}
&\bmath{a}_{i}\equiv\sum_{j=1(\neq i)}^{N}\bmath{a}_{ij}\equiv-\sum_{j=1(\neq i)}^{N}\nabla_{i}\phi(r_{ij}),\\
&\bmath{\varv}_{ij}=\bmath{\varv}_{i}-\bmath{\varv}_{j},
\end{align}
\end{subequations}
and $\bmath{\sigma}_{ij}$ is the normal surface vector perpendicular to the surface of the Landau sphere spanned by the radial vector $\triangle(\bmath{r}_i-\bmath{r}_j)/r_{ij}$ around the position $\bmath{r}_j$ and the surface integral $\oiint$ is taken over the surface components $\text{d}\bmath{\sigma}_{ij}$. The L.H.S of equation \eqref{Eq.BBGKY_truncated} is the same as a standard BBGKY hierarchy except for the truncated DF, while the two terms on the R.H.S appears due to the effects of stars entering or leaving the surface of the Landau sphere; those two extra terms may turn into collisional terms \cite{Cercignani_1972}. The last term vanishes if close encounters are elastic, which is the basic assumption in the present paper while the first line of the R.H.S of equation \eqref{Eq.BBGKY_truncated} corresponds to the Boltzmann collision term. The order of the collision term is estimated as $\sim 1$ if one assumes that relative speed between stars is order of the speed dispersion. 

For the weakly-coupled DF, the contributions from the surface integral vanish; the two terms on the R.H.S of equation \eqref{Eq.BBGKY_truncated} vanish since any star does no exist inside the Landau sphere, i.e. equations \eqref{Eq.weak_coupling} and \eqref{Eq.WC_r12} are valid. Hence, the BBGKY hierarchy for the weakly-coupled DF is
\begin{align}
&\partial_{t}f^{\blacktriangle}_{s}+\sum_{i=1}^{s}\left[\bmath{\varv}_{i}\cdot\nabla_{i}+\sum_{\quad{j=1(\neq i)}}^{s}\bmath{a}_{ij}\cdot\bmath{\partial}_{i}\right]f^{\blacktriangle}_{s}\nonumber\\
&\qquad +\sum_{i=1}^{s}\left[\bmath{\partial}_{i}\cdot\int_{\Omega_{s+1,s+1}}f^{\blacktriangle}_{s+1}\bmath{a}_{i,s+1}\text{d}_{s+1}\right]=0.\label{Eq.BBGKY_WC}
\end{align}
Use of the weakly-coupled DF means that the caveats is considered following the discussion in the present section; the weakly-coupled DF misses counting the effect of due to a few of stars traveling in the Landau sphere inside and the effect of strong encounters are neglected. They come into play as the same order as the $\partial_{t}f(1,t)\sim1$. 
Use of the BBGKY hierarchy, equation \ref{Eq.BBGKY_WC_approx}, for DFs is limited under the evaluation as follows
\begin{enumerate}
	\item rough approximation      $\sim1/N^{3/2}$
	\item no strong encounter      $\sim1/N$
	\item use of weakly-coupled DF $\sim1/N$
	\item use of Heaviside function       $\sim1/N^{2}$
\end{enumerate}
Equation \eqref{Eq.BBGKY_WC_approx} is limited only through the domain of the integrals, not DF themselves. A correct interpretation of the equation is that \emph{the relaxation process in evolution of a star cluster may be considered due to two-body encounters via truncated Newtonian acceleration till the mean density of the system reaches as high as order of $\sim N\bar{n}_{o}$} at which the standard DFs can not be employed in place of the weakly-coupled DF.

Lastly, in limit of $\triangle\rightarrow 0$, one can retrieve a standard BBGKY hierarchy for standard DF from both equations  \eqref{Eq.BBGKY_truncated} and \eqref{Eq.BBGKY_WC}
\begin{align}
&\partial_{t}f_{s}+\sum_{i=1}^{s}\left[\bmath{\varv}_{i}\cdot\nabla_{i}+\sum_{\quad{j=1(\neq i)}}^{s}\bmath{a}_{ij}\cdot\bmath{\partial}_{i}\right]f_{s}\nonumber\\
&\qquad +\sum_{i=1}^{s}\left[\bmath{\partial}_{i}\cdot\int f_{s+1}\bmath{a}_{i,s+1}\text{d}_{s+1}\right]=0.\label{Eq.BBGKY_orth}
\end{align}

\section{The generalized Landau equation for the 'weakly-coupled' distribution function of stars}\label{sec:complete_WC}
In the present section, the weakly-coupled DFs (section \ref{subsec:DF}) is employed to derive a kinetic equation to model evolutions of a 'completely weakly-coupled' star cluster in which no star can approach the other stars closer than the Landau radius $r_\text{o}$. In section \ref{sec:grainess_effect} the effects of truncation of phase-space volume on the collision term (relaxation time of the system) and on the m.f. acceleration (Poisson equation) are discussed.

\subsection{Completely weakly-coupled stellar systems}\label{subsec.compl_WC}
Assume that a star cluster at the early stage of evolution may be modeled by weakly-coupled DFs for stars. The first two equations of the hierarchy, equation \eqref{Eq.BBGKY_WC}, for DF $f_{1}(1,t)$ and the first equation of the hierarchy for DF $f_{1}(2,t)$ respectively read
\begin{subequations}
	\begin{align}
	&(\partial_{t}+\bmath{\varv}_{1}\cdot\nabla_{1})f_{1}(1,t)=-\bmath{\partial}_{1}\cdot\int_{\Omega_{2,2}}f_{2}(1,2,t)\bmath{a}_{12}\text{d}_{2},\label{Eq.1stBBGKY_WC}\\
	&\left(\partial_{t}+\bmath{\varv}_{1}\cdot\nabla_{1}+\bmath{\varv}_{2}\cdot\nabla_{2}+\bmath{a}_{12}\cdot\bmath{\partial}_{12}\right)f_{2}^{\blacktriangle}(1,2,t)\nonumber\\
	&\quad\qquad=-\int_{\Omega_{3,3}}\left[\bmath{a}_{1,3}\cdot\bmath{\partial}_{1}+\bmath{a}_{2,3}\cdot\bmath{\partial}_{2}\right]f_{3}^{\blacktriangle}(1,2,3,t)\text{d}_{3},\label{Eq.2nd_BBGKY_WC}\\
	&(\partial_{t}+\bmath{\varv}_{2}\cdot\nabla_{2})f_{1}(2,t)=-\bmath{\partial}_{2}\cdot\int_{r_{23}>\triangle}f_{2}(2,3,t)\bmath{a}_{23}\text{d}_{3},\label{Eq.1stBBGKY_WC_m.f}
	\end{align}
\end{subequations}
where $\bmath{\partial}_{12}=\bmath{\partial}_{1}-\bmath{\partial}_{2}$ and the domains of DFs and the accelerations are defined only at distances $r_{ij}>\triangle$. 
To simplify equations \eqref{Eq.1stBBGKY_WC} and \eqref{Eq.2nd_BBGKY_WC},  and  correlation formulations can be employed. Ignoring the effect of ternary correlation function $T(1,2,3,t)$ (i.e. the effect of three-body interactions, e.g. triple encounters of stars), the single-, double- and triple- DFs may be, in general, rewritten as following Mayer cluster expansion \citep[e.g.][]{Mayer_1940,Green_1956}\footnote{The DFs and correlation functions for stars, in general, may depend on the number $N$ as
	\begin{align}
	&f(1,t), f(2,t), f(3,t)\propto N,\nonumber\\
	&g(1,2,t), g(2,3,t), g(3,1,t)\propto N(N-1),\label{Eq.scale_g(1,2,t)}
	\end{align}
	where the normalisation condition for DFs and correlation functions follows \citep{Liboff_1966}.}
\begin{subequations}
	\begin{align}
	&f_{1}(1,t)\equiv f(1,t),\label{Eq.singleDF}\\
	&f_{2}(1,2,t)\equiv f(1,2,t)= f(1,t)f(2,t)+\left[g(1,2,t)-\frac{f(1,t)f(2,t)}{N}\right],\label{Eq.doubleDF}\\
&f_{3}(1,2,3,t)=f(1,t)f(2,t)f(3,t)\nonumber\\
&\qquad\qquad\quad+ \left(g(1,2,t)-\frac{f(1,t)f(2,t)}{N}\right)f(3,t)\nonumber\\
&\qquad\qquad\qquad+\left(g(2,3,t)-\frac{f(2,t)f(3,t)}{N}\right)f(1,t)\nonumber\\
&\qquad\qquad\qquad\quad+\left(g(3,1,t)-\frac{f(3,t)f(1,t)}{N}\right)f(2,t).\label{Eq.tripleDF_WC}
	\end{align}
\end{subequations}
where the weak-coupling approximation is employed. The important difference of star clusters from classical plasmas and ordinary neutral gases can be characterised by the effect of smallness parameter, $1/N$, in equations \eqref{Eq.doubleDF} and \eqref{Eq.tripleDF_WC}; the parameter is not ignorable for dense star clusters $\left(10^{5}\lesssim N \lesssim 10^{7}\right)$.

By assuming the system is not gravitaitonally-polarizable, one obtains
\begin{subequations}
\begin{align}
&\left(\partial_{t}+\bmath{\varv}_{1}\cdot\nabla_{1}+\bmath{A}_{1}^{(2,2)}\cdot\bmath{\partial}\right)f(1,t)=-\bmath{\partial}_{1}\cdot\int_{\Omega_{2,2}}g(1,2,t)\bmath{a}_{12}\text{d}_{2},\label{Eq.1stBBGKY_Kadrup}\\
&\left(\partial_{t}+\bmath{\varv}_{1}\cdot\nabla_{1}+\bmath{\varv}_{2}\cdot\nabla_{2}+\bmath{A}_{1}^{(3,3)}\cdot\bmath{\partial}_{1}+\bmath{A}_{2}^{(3,3)}\cdot\bmath{\partial}_{2}\right)g^{\blacktriangle}(1,2,t)\nonumber\\
&=\left(-\bmath{a}_{12}\cdot\bmath{\partial}_{12}+\frac{1}{N}\bmath{A}_{1}^{(3,3)}\cdot\bmath{\partial}_{1}+\frac{1}{N}\bmath{A}_{2}^{(3,3)}\cdot\bmath{\partial}_{2}\right)\left[f(1,t)f(2,t)\Theta(r_{12}-\triangle)\right]\nonumber\\
&\quad-\left(1-\frac{1}{N}\right)\left[\bmath{A}_{1}^{(2,2)}-\bmath{A}_{1}^{(3,3)}\right]\cdot\bmath{\partial}_{1}\left[f(1,t)f(2,t)\Theta(r_{12}-\triangle)\right]\nonumber\\
&\quad-\left(1-\frac{1}{N}\right)\left[\bmath{A}_{2}^{(2,2)}-\bmath{A}_{2}^{(3,3)}\right]\cdot\bmath{\partial}_{2}\left[f(1,t)f(2,t)\Theta(r_{12}-\triangle)\right]\nonumber\\
&\quad-\partial_{1}\cdot\left(\int_{\Omega_{3,3}}\bmath{a}_{13}g(1,3,t)\text{d}_{3}-\int_{\Omega_{2,2}}\bmath{a}_{13}g(1,3,t)\text{d}_{3}\right)f(2,t)\nonumber\\
&\quad-\partial_{2}\cdot\left(\int_{\Omega_{3,3}}\bmath{a}_{23}g(2,3,t)\text{d}_{3}-\int_{\Omega_{2,2}}\bmath{a}_{23}g(2,3,t)\text{d}_{3}\right)f(1,t),\label{Eq.2nd_BBGKY_grained_many}
\end{align}
\end{subequations}
where the lowest OoM of the terms are left with $\mathcal{O}(1)$\footnote{One may realise that the lowest order at equation level is $\sim\mathcal{O}(1/N^{2})$ due to the truncated acceleration $\bmath{A}^{(2,2)}/N$ to hold the self-consistency of the kinetic equation.} and the truncated m.f. accelerations are defined as
\begin{subequations}
	\begin{align}
	&\bmath{A}_{i}^{(2,2)}(\bmath{r}_{i},t)=\left[1-\frac{1}{N}\right]\int_{r_{i3}>\triangle}f(3,t)\bmath{a}_{i3}\text{d}_{3},\quad(i,j=1,2)\\
	&\bmath{A}_{i}^{(3,3)}(\bmath{r}_{i},\bmath{r}_{j},t)=\left[1-\frac{1}{N}\right]\int_{\Omega_{3,3}}f(3,t)\bmath{a}_{i3}\text{d}_{3}.
	\end{align}
\end{subequations}
The last six terms on the R.H.S of equation \eqref{Eq.2nd_BBGKY_grained_many} may be simplified, by neglecting the existence of the third star in  two-body encounter between two stars of concern;
\begin{align}
 \Omega_{2,2}\approx\Omega_{3,3}.
\end{align}
This is possible since the truncated phase-space volume of the truncated DF, equation \eqref{Eq.single_truncated_DF}, for the third star contributes to equation \eqref{Eq.2nd_BBGKY_grained_many} only as a margin of error with order of $\mathcal{O}(1/N^{2})$; corresponding to
\begin{align}
\int_{\Omega_{2,2}}\cdot\quad\text{d}_{3}\approx\int_{\Omega_{3,3}}\cdot\quad\text{d}_{3}+\mathcal{O}\left(1/N^{2}\right).
\end{align}
Hence, equation \eqref{Eq.2nd_BBGKY_grained_many} simply reduces to
\begin{align}
&\left(\partial_{t}+\bmath{\varv}_{1}\cdot\nabla_{1}+\bmath{\varv}_{2}\cdot\nabla_{2}+\bmath{A}_{1}^{(2,2)}\cdot\bmath{\partial}_{1}+\bmath{A}_{2}^{(2,2)}\cdot\bmath{\partial}_{2}\right)g(1,2,t)\nonumber\\
&=-\left[\tilde{\bmath{a}}^{\triangle}_{12}\cdot\bmath{\partial}_{1}+\tilde{\bmath{a}}^{\triangle}_{21}\cdot\bmath{\partial}_{2}\right]f(1,t)f(2,t),\label{Eq.2nd_BBGKY_grained_Kandrup}
\end{align}
where 
\begin{subequations}
	\begin{align}
	\tilde{\bmath{a}}^{\triangle}_{12}=\bmath{a}_{12}-\frac{1}{N}\bmath{A}_{1}^{(2,2)},\\
	\tilde{\bmath{a}}^{\triangle}_{21}=\bmath{a}_{21}-\frac{1}{N}\bmath{A}_{2}^{(2,2)}.
	\end{align}
\end{subequations}
Employing the method of characteristics, one obtains the correlation function from equation \eqref{Eq.2nd_BBGKY_grained_Kandrup}
\begin{align}
&g^{\blacktriangle}(1,2,t)\nonumber\\
&=g^{\blacktriangle}(1(t-\tau),2(t-\tau),t-\tau)\nonumber\\
&\quad-\int^{t}_{t-\tau}\left[\tilde{\bmath{a}}^{\triangle}_{12}\cdot\bmath{\partial}_{1}+\tilde{\bmath{a}}^{\triangle}_{21}\cdot\bmath{\partial}_{2}\right]_{t=t'}f\left(1(t'),t'\right)f\left(2(t'),t'\right)\Theta(r_{12}(t')-\triangle)\text{d}t'.\label{Eq.Generalized_Landau_truncated_g}
\end{align}

In the scenario for the g-Landau equation in \citep{Kandrup_1981}, all the stars in a star cluster are perfectly uncorrelated at the beginning of correlation time $t-\tau$, implying that the destructive term $g(1(t-\tau),2(t-\tau),t-\tau)$ vanishes at two-body DF level. To apply the same simplification for a secular evolution of the system of concern, one must necessarily consider the memory effect, that is of importance if the time duration between encounters is comparable to the correlation-time scale. The memory effect, however, may be of less significance in stellar dynamics due to the violent relaxation, short-range two-body encounters, spatial inhomogeneities and anisotropy \citep[e.g.][pg. 34]{Saslaw_1985}. Hence, the destructive term on the R.H.S of equation \eqref{Eq.Generalized_Landau_truncated_g} may vanish. One obtains the g-Landau equation with the effect of discreteness from equations \eqref{Eq.1stBBGKY_Kadrup} and \eqref{Eq.Generalized_Landau_truncated_g}
\begin{align}
&\left(\partial_{t}+\bmath{\varv}_{1}\cdot\nabla_{1}+\bmath{A}_{1}^{(2,2)}\cdot\bmath{\partial}_{1}\right)f(1,t)\nonumber\\
&=\bmath{\partial}_{1}\cdot\int_{\Omega_{2,2}}\text{d}_{2} \bmath{a}_{12}\int_{0}^{\tau}\text{d}\tau'\left[\tilde{\bmath{a}}_{12}^{\triangle}\cdot\bmath{\partial}_{1}+\tilde{\bmath{a}}_{21}^{\triangle}\cdot\bmath{\partial}_{2}\right]_{t-\tau'}\nonumber\\
&\quad\times f(1(t-\tau'),t-\tau')f(2(t-\tau'),t-\tau').\label{Eq.Generalized_Landau_truncated}
\end{align}

The effect of retardation in the collision term of equation \eqref{Eq.Generalized_Landau_truncated} may be discussed. Since the trajectory of test star is chracterised by equation \eqref{Eq.characteristics_m.f.}, the correlation time would be at most the free-fall time of test star under the effect of the m.f. acceleration while the shortest correlation time scale is longer than the time scale for test star to travel across a Landau sphere to hold the weak-coupling approximation;
\begin{align}
\mathcal{O}(1/N)<t_\text{cor}\lesssim\mathcal{O}(1),
\end{align} 
meaning the non-Markovian effect on the relaxation process is less significant;
\begin{align}
\mathcal{O}(1/N^{2})<\frac{t_\text{cor}}{t_\text{rel}}\lesssim\mathcal{O}(1/N).
\end{align} 
Hence, one may assume the Markovian limit\footnote{For the Markovian limit, one should not change the other arguments of the DF in the collision term since the changes in momentum and position of test star in encounter is not ignorable due to the effect of m.f. acceleration.} for the collision term for the correlation time $0<\tau'<t_\text{cor}$
\begin{align}
f(1(t-\tau'),t-\tau')f(2(t-\tau'),t-\tau')\approx f(1(t-\tau'),t)f(2(t-\tau'),t),
\end{align}
Taking the limit of $\tau\to\infty$, one obtains
\begin{align}
&\left(\partial_{t}+\bmath{\varv}_{1}\cdot\nabla_{1}+\bmath{A}_{1}^{(2,2)}\cdot\bmath{\partial}_{1}\right)f(1,t)\nonumber\\
&=\bmath{\partial}_{1}\cdot\int_{\Omega_{2,2}}\text{d}_{2} \bmath{a}_{12}\int_{0}^{\infty}\text{d}\tau'\nonumber\\
&\quad\times\left[\tilde{\bmath{a}}_{12}^{\triangle}\cdot\bmath{\partial}_{1}+\tilde{\bmath{a}}_{21}^{\triangle}\cdot\bmath{\partial}_{2}\right]_{t-\tau'} f(1(t-\tau'),t)f(2(t-\tau'),t).\label{Eq.Generalized_Landau_truncated_Marko}
\end{align}
Employing the anti-normalization condition, equation \eqref{Eq.Anti-norm}, for the correlation function and taking a limit of $\triangle\to 0$, one may retrieve the \citep{Kandrup_1981}'s g-Landau equation;
\begin{align}
&\left(\partial_{t}+\bmath{\varv}_{1}\cdot\nabla_{1}+\bmath{A}_{1}\cdot\bmath{\partial}_{1}\right)f(1,t)=\bmath{\partial}_{1}\cdot\int\text{d}_{2} \tilde{\bmath{a}}_{12}\int_{0}^{\infty}\text{d}\tau'\nonumber\\
&\quad\times\left[\tilde{\bmath{a}}_{12}\cdot\bmath{\partial}_{1}+\tilde{\bmath{a}}_{21}\cdot\bmath{\partial}_{2}\right]_{t-\tau'} f(1(t-\tau'),t)f(2(t-\tau'),t),\label{Eq.Generalized_Landau}
\end{align}
where the statistical acceleration can be found in the forms
\begin{subequations}
	\begin{align}
	\tilde{\bmath{a}}_{12}=\bmath{a}_{12}-\frac{1}{N}\bmath{A}_{1},\\
	\tilde{\bmath{a}}_{21}=\bmath{a}_{21}-\frac{1}{N}\bmath{A}_{2}.
	\end{align}
\end{subequations}
The truncated g-Landau equation \eqref{Eq.Generalized_Landau_truncated} is different from the g-Landau equation \eqref{Eq.Generalized_Landau}, not only in the domain of interaction range, but also in the form of physical quantities; the truncated- acceleration and collision term. The truncation of the phase-space volume in integrals is termed as 'the effect of discreteness' in the present work and discussed in section \ref{sec:grainess_effect}.

\subsection{g-Landau equation with truncated pair-wise potential}
The present section derives the g-Landau equation with completely weakly-coupled DF. To do so, one needs to modify the g-Landau equation for weakly-coupled DF. One can convert the m.f. acceleration into 

Also, m.f. accelerations coincides due to the condition for truncation
\begin{align}
\bmath{A}^{(a)}(\bmath{r}_{1},t)=\bmath{A}^{(|phi)}(\bmath{r}_{1},t)-\triangle^{2}a_{1}(r_{1},\triangle)=\bmath{A}^{(|phi)}(\bmath{r}_{1},t)
\end{align}
The collision term also must be converted

The g-Landau equation with weakly-couple DF and Poisson equation reduce to
\begin{align}
&\left(\partial_{t}+\bmath{\varv}_{1}\cdot\nabla_{1}+\bmath{A}^{(\Phi)}(\bmath{r}_{1},t)\cdot\bmath{\partial}_{1}\right)f(1,t)\nonumber\\
&=\bmath{\partial}_{1}\cdot\int_{\Omega_{2,2}}\text{d}_{2} \bmath{a}_{12}\int_{0}^{\infty}\text{d}\tau'\nonumber\\
&\quad\times\left[\tilde{\bmath{a}}_{12}^{\triangle}\cdot\bmath{\partial}_{1}+\tilde{\bmath{a}}_{21}^{\triangle}\cdot\bmath{\partial}_{2}\right]_{t-\tau'} f(1(t-\tau'),t)f(2(t-\tau'),t).\label{Eq.Generalized_Landau_truncated_Marko_final}\\
&\nabla^{2}_{1}\Phi^{(2,2)}(\bmath{r}_{1},t)=\nabla_{1}\bmath{A}^{(\Phi)}(\bmath{r}_{1},t)=4\pi G \rho^{(\Phi)}(\bmath{r}_{1},t).\\
&\rho^{(\Phi)}(\bmath{r}_{1},t)=Gm \left(1-\frac{1}{N}\right)\int (1-\triangle\hat{r}\cdot\nabla_{1}) n_{1}(\bmath{r}_{1}-\triangle\hat{r})\text{d}\Omega
\end{align}

\section{The relaxation time and statistical acceleration}\label{sec:grainess_effect_tr}
In the present section, the effects of 'weakly-coupled' DF on the relaxation time of a system modeled by equation \eqref{Eq.Generalized_Landau_truncated_Marko} under homogeneous- and local- approximations

\subsection{The effect of discretness on the relaxation time}
To evaluate the effect of 'discreteness' on the relaxation time, assume test star follows the rectilinear motion, equation \eqref{Eq.characteristics_rectilinear}, and the encounter is local for the truncated g-Landau collision term in equation \eqref{Eq.Generalized_Landau_truncated_Marko}, meaning the truncated Landau collision term is examined;
\begin{align}
I_\text{L}^{\blacktriangle}=&\bmath{\partial}_{1}\cdot\int_{\Omega_{2,2}} \bmath{a}_{12}\int_{0}^{\infty}\left[\bmath{a}_{12}\left(t-\tau'\right)\right]_{r_{12}\left(t-\tau'\right)>\triangle}\text{d}\tau'\text{d}^{3}\bmath{r}_{12}\nonumber\\
                         &\qquad\cdot\bmath{\partial}_{12} f(1,t)f(\bmath{r}_{1},\bmath{p}_{2},t)\text{d}^{3}\bmath{p}_{2},\label{Eq._Landau_truncated_Marko}
\end{align}
where the effect of non-ideality (retardation and spatial non-locality) for the Landau collision term was neglected for simplicity. The Fourier-transform of the acceleration of star 1 due to star 2 at distances $r_{12}>\triangle$ is as follows\footnote{It is to be noted that the Fourier transform of the potential $\phi_{12}$ typically done to find the explicit form of the Landau collision term necessitates a 'convergent factor',$e^{-\lambda r_{12}}$, where $\lambda$ is a vanishing low number to be taken as zero after the Fourier transform. The factor can remove singularities of (generalised) functions on complex planes and slow decays of potentials in three dimensional spaces \citep[e.g][]{Adkins_2013}. One, however, does not need to employ the factor in the Fourier transform of the truncated acceleration, $\bmath{a}_{12}\Theta(r_{12}>\triangle)$, and even in the corresponding inverse Fourier transform, $\mathcal{F^{-1}}\left[\mathcal{F}\left[\bmath{a}_{12}\Theta(r_{12}>\triangle)\right]\right]$. Rendering the transform, $\mathcal{F^{-1}}\left[\mathcal{F}\left[\bmath{a}_{12}\Theta(r_{12}>\triangle)\right]\right]$, is a simple task, hence it will be left for readers; one will need the following identity to find the step function
\begin{align}
\int^{\infty}_{0}\frac{\sin k}{k}\text{d}k=\frac{\pi}{2}.
\end{align}
}
\begin{subequations}
	\begin{align}
	\mathcal{F}\left[\bmath{a}_{12}(r_{12}>\triangle)\right]&=-Gm\int_{\Omega_{2,2}}\exp(-\text{i}\bmath{k}\cdot \bmath{r}_{12})\frac{\bmath{r}_{12}}{r_{12}^{3}}\text{d}^{3}\bmath{r}_{12},\\
	&=-\frac{Gm\sin[k\triangle]}{2i\pi^{2}k^{2}\triangle}\hat{k}. \label{Eq.Fourier_a}
	\end{align}
\end{subequations}
The same transform must be employed for $\bmath{a}_{12}(t-\tau')$ in equation \eqref{Eq._Landau_truncated_Marko} but the time of $\bmath{r}$ is fixed to $t-\tau'$ and the corresponding wavenumber must be exploited. It is to be noted that equation \eqref{Eq.Fourier_a} is in essence the same as the Fourier transform of the truncated acceleration $\bmath{a}_{12}\Theta(r_{12}-\triangle)$, meaning the corresponding acceleration of star 1 is null within the volume of the Landau sphere. This is since the existence of stars in the Landau sphere is not of concern due to the spatial locality and the effect of truncation on DF must be controlled through truncation of acceleration. In the limit of $\triangle\to 0$, equation \eqref{Eq.Fourier_a} results in a well-known Fourier transform of acceleration or pair-wise Newtonian potential in wave kinetic theory \citep[e.g.][Appendix C]{Chavanis_2012}
\begin{subequations}
	\begin{align}
	\lim_{\triangle\to 0}\mathcal{F}\left[\bmath{a}_{12}\right]&=-\frac{Gm}{2i\pi^{2}k}\hat{k},\\
	&\xrightarrow[\times\frac{1}{-\text{i}\bmath{k}}]{}\mathcal{F}\left[\phi_{12}\right]\hat{k}.
	\end{align}
\end{subequations}
After a proper calculation following \citep[][Appendix C]{Chavanis_2012}, the collisional term results in
\begin{subequations}
	\begin{align}
	&I_\text{L}^{\blacktriangle}=\bmath{\partial}_{1}\cdot \int \overleftrightarrow{T}(\bmath{p}_{12},\bmath{p}_{2})\cdot\bmath{\partial}_{12} f(\bmath{r}_{1},\bmath{p}_{1},t)f(\bmath{r}_{1},\bmath{p}_{2},t) \text{d}^{3}\bmath{p}_{2}, \\
	&\overleftrightarrow{T} \equiv -B\frac{\bmath{p}^{2}_{12}\overleftrightarrow{I}-\bmath{p}_{12}\bmath{p}_{12}}{p^{3}},\qquad (\bmath{p}_{12}\equiv\bmath{p}_{1}-\bmath{p}_{2})\\
	&B\equiv 2\pi G m^{2}\int^{\infty}_{0}\frac{\sin^{2}[k\triangle]}{k^{3}\triangle^{2}}\text{d}k. \label{Eq.g_Logarithm}
	\end{align}
\end{subequations}
where for expressions of the tensor $\overleftrightarrow{T}$, typical dyadics are exploited. Following the works \citep{Severne_1976,Kandrup_1981,Chavanis_2013a} if one assumes the cut-offs $k\in[2\pi/R,2\pi/r_\text{o}]$ on each limit of the integral domain of the collision term, the factor $B$, equation \eqref{Eq.g_Logarithm}, explicitly reads
\begin{align}
B=\ln[N]+1.5+\sum_{m=1}^{\infty}\frac{(-16\pi^{2})^{m}}{2m(2m)!}+\mathcal{O}(1/N^{2}),\label{Eq.B}
\end{align}
where the following indefinite integral formula \citep[e.g.][]{Zeidler_2004} was employed\footnote{A Similar calculation for a weakly-nonideal self-gravitating system appears in \citep{Bose_2012}, in which the upper limit is also assigned on the domain of the integration.} 
\begin{align}
	\int\frac{\cos[\alpha k]}{k}\text{d}k=\ln[\alpha k]+\sum_{m=1}^{\infty}\frac{\left(-[\alpha k]^{2}\right)^{m}}{2m(2m)!}.\label{Eq.math_formula}
\end{align}
It would be obvious that the lower limit of the distance $r_{12}$, the Landau radius, can not remove the logarithmic singularity in the collision term as shown in equation \eqref{Eq.B}. This is of course since an application of the weak-coupling approximation to the g-Landau collision term is inconsistent especially at $r_{12}\to r_\text{o}$;to avoid the singularity associated with high wavenumbers, one needs all the higher orders of weak-coupling approximation as correction to the rectilinear-motion approximation, or the trajectory  of test star must follow pure Newtonian two-body problem, equation \eqref{Eq.characteristics_two_body}. Since the value of the parameter $c$ in equation \eqref{Eq.ratio_R_ro} is in essence a user-choice parameter, the following ideal (often-employed in wave kinetic theory) relation is assumed for simplicity
\begin{align}
R=Nr_\text{o}.\label{Eq.simple_R_ro}
\end{align}

\begin{table}
	\caption{The decrease rate of the Coulomb logarithm, $100(1-B/\ln[N])$, due to the effect of discreteness.}
	\label{tabel:trunc_Coulomb_log}
	\begin{tabular}{|l|c|l|c|}
		\hline
		$N$                          & decrease rate [\%]   & $N$                          & decrease rate [\%] \\ 
		\hline
		$10^{5}$                     & 14.02                   & $10^{9}$                     & 7.790 \\
		$10^{6}$                     & 11.69                   & $10^{10}$                    & 7.011 \\
		$10^{7}$                     & 10.02                   & $10^{11}$                    & 6.373 \\
		$10^{8}$                     & 8.764                   &                              &        \\
		\hline
	\end{tabular}\\
\end{table}

The result of numerical integration of the factor $B$, equation \eqref{Eq.g_Logarithm}, is as follows
\begin{align}
B-\ln[N]=1.5+ \sum_{m=1}^{\infty}\frac{\left(-[4\pi]^{2}\right)^{m}}{2m(2m)!}\approx-3.11435,
\end{align}
and the decrease rate of the Coulomb logarithm for different $N$ is shown in Table \ref{tabel:trunc_Coulomb_log}. It turns out, the effect of discreteness \emph{decreases} the coulomb logarithm $\ln[N]$ for relatively low-number star cluster $\left(N=10^{5}\right)$ by 14.0 $\%$, for high-number cluster $\left(N=10^{7}\right)$ by 10.0 $\%$, and (as a reference) for large galaxies $\left(N=10^{11}\right)$ by 6.37$\%$; accordingly, the corresponding relaxation times \emph{increase} from typical one (that has the same physical condition but effect of discreteness) by the same factors. 

This result clearly concerns practitioners since the modification of the Coulomb logarithm is relatively large. For example, the Coulomb logarithm was originally underestimated by Chandrasekhar under use of neighboring-encounter approximation \citep{Chandra_1943a}, meaning the order of coulomb logarithm is approximately modified by
\begin{align}
\eta\equiv 100\left(1-\frac{\ln[N^{2/3}]}{\ln[N]}\right)=33 \%.
\end{align} 
The truncated DF pushes back the logarithm to the classical value.

\subsection{relaxation time with upper limit}
The result of the discreteness encourages one heuristically assign the upper bound on the interaction range too. After applying the Fourier transform of acceleration limited on $[r_\text{o}<r_{12}<R]$
\begin{align}
&\mathcal{F}\left[\bmath{a}_{12}(R>r_{12}>\triangle)\right]&=\frac{Gm}{2i\pi^{2}}\left(\frac{\sin[kR]}{k^{2}R}-\frac{\sin[k\triangle]}{k^{2}\triangle}\right)\hat{k}. \label{Eq.Fourier_a_up}
\end{align}
one obtains the new constant $B'$ instead of $B$, equation \eqref{Eq.B},
\begin{align}
B'\equiv 2\pi G m^{2}\int^{\infty}_{0}\frac{1}{k}\left(\frac{\sin[kR]}{kR}-\frac{\sin[k\triangle]}{k\triangle}\right)^{2}\text{d}k. \label{Eq.g_Logarithm_up}
\end{align}
The new terms can be calculated as follows
\begin{align}
B'&=B-\gamma-\ln[4\pi]-\sum_{m=1}^{\infty}\frac{(-16\pi^{2})^{m}}{2m(2m)!}+\mathcal{O}(1/N)\\
&=\ln[N]+1.5-\gamma-\ln[4\pi]+\mathcal{O}(1/N). \label{Eq.g_Logarithm_up}
\end{align}
where $\gamma$ is Euler-Mascheroni constant and the value is $\gamma\approx0.5772$ .To find the third equality \eqref{Eq.B} is employed and to find the second the following identity (e.g. Grad Reby equation 3.761) is employed 
The explicit form of the constant $B'$ is
\begin{align}
\int^{\infty}_{0}\frac{\sin(ax)}{x^{2}}\text{d}x=\sin(a)+a\left(\gamma+\ln[4\pi]+\sum_{m=1}^{\infty}\frac{(-16\pi^{2})^{m}}{2m(2m)!}\right) ,\label{Eq.identity_2}
\end{align}
As expected the new Coulomb logarithm $B'$ is closer to the Coulomb logarithm than the $B$ after including the upper cut-off at $r_{12}=R$.
\begin{align}
B'-\ln[N]\approx-1.608.
\end{align}
The value of the $B'$ is relatively close to the value obtained in Chandra1941 i.e. $\ln[N]-0.2367$ where the homogeneous background and nearest-neighboring approximation are taken for Holtsmark distribution of force fields. In case of classical plasma Fokker-Planck for the $\ln[N]-0.4420$. Those values are obtained by assuming the background takes a Maxwellian while the value of $B'$ is purely due to the nature of Newtonian force and cut-off on the spaces. One must, of course, employ the g-landau kinetic equation itself to find out the correct modification to the Coulomb logarithm due to inhomogeneity though, the above discussion well describes the effect on the logarithm of long-range nature of Newtonian force and finiteness of the system. 

Under the assumption of local encounter, limiting the range of encounter distance between the Landau radius and system size results in `dominat effect' in stochastic theory, while that results in `non-dominant effect' based on  BBGKY hierarchy. The former is of significance at relaxation time scale while the latter is at secular time scale.

\section{Truncated Poisson equations and core-halo structure}\label{sec:grainess_effect_Poisson}
The applicability of the truncated acceleration and potential to core-collapse problem is discussed. As discussed in section \ref{sec:truncated_BBGKY}, they are meaningful till the mean density reaches order of $N^{2}\bar{n}$, which is high enough to see the self-similar evolution. Hence, a curiosity in the present section is how no acceleration or no potential in the Landau sphere affects the m.f. acceleration and potential.  Poisson equations for the m.f.- acceleration and potential are derived in section \ref{subsec.Poisson}, and the truncated Poisson equations are applied a toy model for core-halo structure of a spherically symmetric cluster in sections \ref{subsec:trunc_Aa} and \ref{subsec:trunc_Aphi}.

\subsection{Poisson equation for the truncated density}\label{subsec.Poisson}
Utilizing the identity
\begin{align}
\nabla^{2}_{1}\left(\frac{1}{r_{12}}\right)=0 \qquad (r_{12}>\triangle),
\end{align}
one can derive Poisson equation for the truncated m.f. acceleration (or the 'truncated Poisson equation')
\begin{align}
\nabla_{1}\cdot\bmath{A}_{1}^{(2,2)}=-Gm\left(1-\frac{1}{N}\right) \int n_{1}(\bmath{r}_{1}-\triangle\hat{r})\text{d}\Omega,\label{Eq.grained_poisson_A}
\end{align}
where $\text{d}\Omega$ is the element of solid angle spanned by a unit vector $\hat{r}$ in radial direction. Typical observations for star clusters are done at radii 0.01 $\sim$ 1 pc from the center of the clusters even for possibly collapsed clusters \citep[e.g.][]{King_1985,Lugger_1995}; this corresponds with $r_{1}>> r_\text{o}$ if the system dimension reaches $\sim$ tens of parsec. Hence, as a practical application, one may approximate the truncated Poisson equation \eqref{Eq.grained_poisson_A} to
\begin{align}
\nabla_{1}\cdot\bmath{A}_{1}^{(2,2)}\approx& -4\pi Gm\left(1-\frac{1}{N}\right) n_{1}(\bmath{r}_{1})+\triangle Gm\int\hat{r}\cdot\nabla_{1}n_{1}(\bmath{r}_{1})\text{d}\Omega\nonumber\\
&\qquad+\mathcal{O}(1/N^{2}),\\
=&-4\pi Gm \left(1-\frac{1}{N}\right)n_{1}(\bmath{r}_{1})+\mathcal{O}(1/N^{2}).
\end{align}
The relation of the truncated m.f. acceleration with the truncated potential may be written as
\begin{align}
\nabla^{2}_{1}\Phi^{(2,2)}=-\nabla_{1}\cdot\bmath{A}^{(2,2)}_{1}-Gm\left(1-\frac{1}{N}\right) \int\triangle\hat{r}\cdot\nabla_{1}n_{1}(\bmath{r}_{1}-\triangle\hat{r})\text{d}\Omega.
\end{align}
Employing equation \eqref{Eq.grained_poisson_A}, the Poisson equation for the truncated potential reads
\begin{align}
\nabla^{2}_{1}\Phi^{(2,2)}=Gm \left(1-\frac{1}{N}\right)\int (1-\triangle\hat{r}\cdot\nabla_{1}) n_{1}(\bmath{r}_{1}-\triangle\hat{r})\text{d}\Omega,\label{Eq.grained_Poisson_Phi}
\end{align}
and in a limit of $\triangle\to 0$
\begin{align}
\nabla^{2}_{1}\Phi^{(2,2)}\approx 4\pi Gm \left(1-\frac{1}{N}\right)n_{1}(\bmath{r}_{1})+Gm \int\left[\triangle\hat{r}\cdot\nabla_{1}\right]^{2}n_{1}(\bmath{r}_{1})\text{d}\Omega.\label{Eq.grained_Poisson_Phi_approx}
\end{align}
The standard Poisson equation for the m.f. potential of star clusters is also applicable to any star cluster at radii $r_{1}>>r_\text{o}$ since the second term on the R.H.S in equation \eqref{Eq.grained_Poisson_Phi_approx} is order of $\mathcal{O}\left(1/N^{2}\right)$. It is to be noted that the truncated Poisson equation \eqref{Eq.grained_poisson_A} or \eqref{Eq.grained_Poisson_Phi} itself shows a kind of coarse-graining on the surface of the Landau sphere through istropising the density of the system of concern at radius of $\triangle$. 
 
For theoretical/numerical studies of stellar dynamics, the dynamics inside the Landau sphere may be of importance since the core size of the system of concern can mathematically reach the size of the Landau radius and the halo may have a strong inhomogeneity in density as a result of gravothemal-instability \citep[e.g][]{Cohn_1979,Takahashi_1995}. In this case, one can no longer employ typical Poisson equation, hence one must hold the form of the truncated Poisson equation \eqref{Eq.grained_poisson_A} or \eqref{Eq.grained_Poisson_Phi}.  For application purpose one can rewrite the truncated Poisson equations for spherically symmetric system as follows

\subsection{The effect of truncated pair-wise acceleration on density profile and m.f. acceleration}\label{subsec:trunc_Aa}
To consider the effect of truncation of the pair-wise acceleration, one may employ the following angle-averaged density
\begin{align} 
\bar{n}^{(a)}(\bmath{r}_{1},t)=\frac{1}{4\pi}\int n(\bmath{r}_{1}-\triangle \hat{r}',t)\text{d}\Omega'
\end{align}
where the superscript $(a)$ in the density means that the coarse-graining of the density originates from truncation of pair-wise acceleration, or DFs. If one assumes that the system of concern is a spherically symmetric $n=n(r_{1})$, as explained in Appendix \ref{Appendix:density_a12}, one can find the following reduced form of the averaged density and Poisson equation
\begin{subequations}
\begin{align}
&\frac{1}{r^{2}}\frac{\text{d}}{\text{d}r}\left[r^{2}A^{(a)}(r)\right]=\bar{n}^{(a)}(r,t)\label{Eq.Poisson_spheric_a12}\\
&\bar{n}^{(a)}(r,t)=\frac{1}{2r\triangle}\int^{r+\triangle}_{r_{>}-r_{<}}n(r',t)r'\text{d}r'\label{Eq.nbar_formula}
\end{align}
\end{subequations}

where, for brevity, the displacement vector $\bmath{r}_{1}$ and m.f. acceleration $\bmath{A}^{(2,2)}_{1}(\bmath{r}_{1},t)$ for star 1 are relplaced by $\bmath{r}$ and $\bmath{A}^{(a)}(\bmath{r},t)$ respectively and use of the following notations is made  $r_{>}=\max{(r,\triangle)}$ and $r_{<}=\min{(r,\triangle)}$.  Since the truncated DF is relevant to fine structure around the core on scale of $\sim R/N$ i.e. core collapse at the self-similar regime in evolution of the cluster, one may consider the following density profile as a toy model for self-similar evolution of a cluster
\begin{align}
n_\text{ch}(r)=\frac{n_\text{o}}{\left(1+\left[\frac{r}{r_\text{o}}\right]^{2}\right)^{\alpha/2}}
\end{align}
where $\alpha\approx 2.23$. The modified Hubble profile may not be suitable to a modeling of dense clusters though, it still can be characterized by `core radius' $r_\text{o}$ and halo-density profile $\frac{\text{d}n_\text{ch}(r)}{\text{d}r}\approx -2.23$. Especially, the profile provides one the following analytical form of the coarse-grained density
 \begin{align}
\bar{n}_\text{ch}^{(a)}(\bmath{r},t)=\frac{n_\text{o}r_\text{o}^{2}}{4 r_{1}\triangle}\frac{1}{1-\alpha/2}\left(\frac{1}{\left(1+\left[\frac{r+\triangle}{r_\text{o}}\right]^{2}\right)^{\beta}}-\frac{1}{\left(1+\left[\frac{r-\triangle}{r_\text{o}}\right]^{2}\right)^{\beta}}\right)
\end{align}
where $\beta=\alpha/2-1$. In the present work, the core size is assumued $r_\text{o}=1\times10^{-6}$ compared to the system size $R=1$. Hence, one can find the relation $R\sim Nr_\text{o}\sim N\triangle$ which can be easily achieved by numerical studies. The coarse-grained density $n_\text{ch}(r)$ coincides with the asymptote of density $n_\text{ch}(r)$ on large scale ($r\to \infty$) while it lowers density due to the averaging process on small scales as follows
\begin{align}
\bar{n}_\text{ch}(r)=\begin{cases}
n_\text{o}\left(\frac{r_\text{o}}{r}\right)^{\alpha}\qquad\qquad (r\to \infty)\\
\frac{n_\text{o}}{\left(1+\left[\frac{\triangle}{r_\text{o}}\right]^{2}\right)^{\beta+1}}\quad\qquad (r\to 0)
\end{cases}
\end{align}
The corresponding asymptotes of the truncated acceleration take
\begin{align}
A_\text{ch}^{(a)}(r)=\begin{cases}-\left(\frac{r_\text{o}}{r}\right)^{\alpha}r\qquad\qquad (r\to \infty)\\
                                                     -\frac{r}{3}\frac{n_\text{o}}{\left(1+\delta^{2}\right)^{\alpha/2}}\quad\qquad (r\to 0)\end{cases}
\end{align}

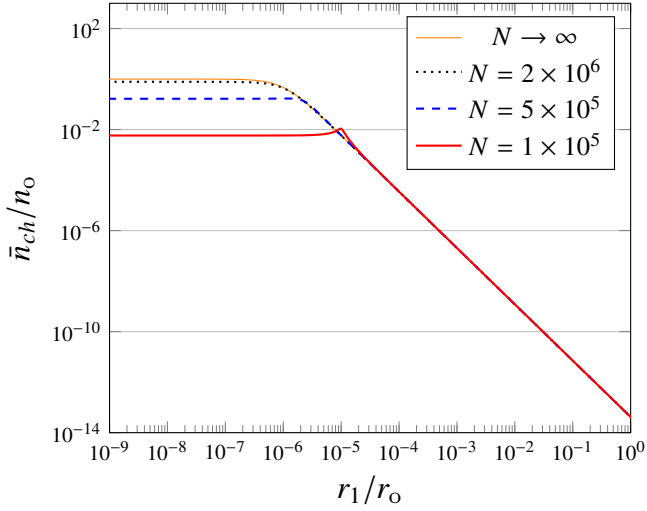
\begin{figure}
	\centering
	\begin{tikzpicture}[scale=1]
	\begin{loglogaxis}[ xlabel=\Large{$r_{1}/r_\text{o}$},ylabel=\Large{$\bar{n}_{ch}/n_\text{o}$},xmin=1e-9,xmax=1,ymin=1e-14,ymax=1e3, legend pos=north east,   ymajorgrids, yminorgrids, minor grid style={orange,dashed} ]
	\addplot [color = orange,mark=no,thin, solid] table[x index=0, y index=1]{rho.txt}; 
	\addlegendentry{\large{$N\to\infty$}};
	\addplot [color = black, thick, dotted] table [x index=0, y index=2]{rho.txt};
	\addlegendentry{\large{$N=2\times10^{6}$}};
	\addplot [color = blue, thick, dashed] table [x index=0, y index=3]{rho.txt};
	\addlegendentry{\large{$N=5\times10^{5}$}};
	\addplot [color = red, thick, solid] table [x index=0, y index=4]{rho.txt};
	\addlegendentry{\large{$N =1\times10^{5}$}};
	\end{loglogaxis}
	\end{tikzpicture}
	\caption{Density profile of a toy model for different value of $\delta(\equiv\triangle/r_\text{o})$  }
	\label{fig:density}
\end{figure}

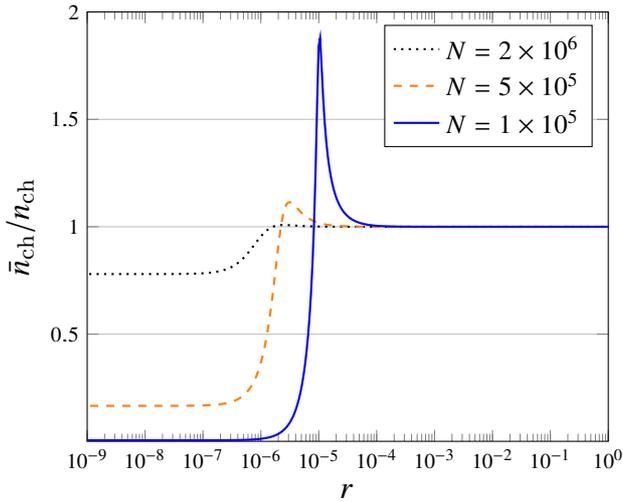
\begin{figure}
	\centering
	\begin{tikzpicture}[scale=1]
	\begin{semilogxaxis}[ xlabel=\Large{$r$},ylabel=\Large{$\bar{n}_\text{ch}/n_\text{ch}$},xmin=1e-9,xmax=1,ymin=1e-3,ymax=2, legend pos=north east,   ymajorgrids]
	\addplot [color = black,mark=no,thick, dotted] table[x index=0, y index=1]{ratio_rho.txt}; 
	\addlegendentry{\large{$N =2\times10^{6}$}};
	\addplot [color = orange, thick, dashed] table [x index=0, y index=2]{ratio_rho.txt};
	\addlegendentry{\large{$N =5\times10^{5}$}};
	\addplot [color = blue, thick, solid] table [x index=0, y index=3]{ratio_rho.txt};
	\addlegendentry{\large{$N =1\times10^{5}$}};
	\end{semilogxaxis}
	\end{tikzpicture}
	\caption{The ratio of coarse-grained density profile $\bar{n}_\text{ch}(r)$ to the fine-grained density $n_\text{ch}(r)$ for different value of $\delta(\equiv\triangle/r_\text{o})$  }
	\label{fig:ratio_density}
\end{figure} 

Figure \ref{fig:density} compares the density profiles of the coarse-grained $\bar{n}_\text{ch}$ for different values of $\delta(\equiv \triangle/r_{o})$. When the size of the Landau radius $\triangle$ is less than or close to the core radius $r_\text{o}$ ($\delta\sim1$)the density depletes in the core. If the core collapse developed well i.e. the size core is smaller than the landau distance ($r_\text{o}<\triangle$), the density profile spikes at $r\sim \triangle$ due to the assumption that two star can not approach closer than Landau radius. This can be well seen in figure \ref{fig:ratio_density} in which the ratio of the coarse-grained density $\bar{n}_\text{ch}(r)$ to raw density $n(r)$ is taken, while the effect of discreteness does not affect the halo structure as expected. 

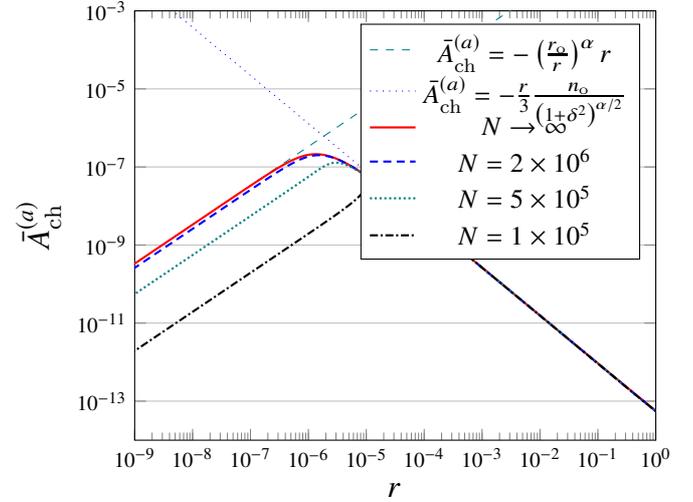
\begin{figure}
	\centering
	\begin{tikzpicture}[scale=1]
	\begin{loglogaxis}[ xlabel=\Large{$r$},ylabel=\Large{$\bar{A}^{(a)}_\text{ch}$},xmin=1e-9,xmax=1,ymin=1e-14,ymax=1e-3, legend pos=north east,   ymajorgrids, yminorgrids, minor grid style={orange,dashed} ]
	\addplot [color =teal,mark=circle,thin, dashed] table[x index=0, y index=1]{Ao.txt}; 
	\addlegendentry{\large{$\bar{A}^{(a)}_\text{ch}=-\left(\frac{r_\text{o}}{r}\right)^{\alpha}r $}};
	\addplot [color = blue,mark=no,thin, dotted] table[x index=0, y index=2]{Ao.txt}; 
	\addlegendentry{\large{$\bar{A}^{(a)}_\text{ch}=-\frac{r}{3}\frac{n_\text{o}}{\left(1+\delta^{2}\right)^{\alpha/2}}$}};
      	\addplot [color = red,mark=no,thick, solid] table[x index=0, y index=3]{Ao.txt};
	\addlegendentry{\large{$N\to\infty$}};
	\addplot [color = blue,mark=no,thick, densely dashed] table[x index=0, y index=1]{A05.txt};
	\addlegendentry{\large{$N =2\times10^{6}$}};
	\addplot [color = teal,mark=no,thick, densely dotted] table[x index=0, y index=1]{A2.txt};
	\addlegendentry{\large{$N =5 \times10^{5}$}};
	\addplot [color = black,mark=no,thick, densely dashdotted] table[x index=0, y index=1]{A10.txt};
	\addlegendentry{\large{$N =1\times10^{5}$}};
	\end{loglogaxis}
	\end{tikzpicture}
	\caption{mean field acceleration of stars due to the effect of the truncation on pair-wise acceleration}
	\label{fig:A_a}
\end{figure}

To see effect of discreteness on the truncated m.f. acceleration, one needs to numerically integrating Poisson equaiton \eqref{Eq.Poisson_spheric_a12} (with B.C. $A^{(a)}(r=0)=0$). To do so, normalize the Poisson equation for static spherically symmetric system as follows
\begin{subequations}
\begin{align}
&\frac{1}{\xi^{2}}\frac{\text{d}}{\text{d}\xi}\left[\xi^{2}\tilde{A}^{(a)}(\xi)\right]=\frac{\bar{n}^{(a)}(\xi)}{n_\text{o}}\label{Eq.Poisson_spheric_a12}\\
&\tilde{A}^{(a)}=4\pi GmRn_\text{o}A^{(a)}\\
&\xi=R\tilde{r}
\end{align}
\end{subequations}
Figure \ref{fig:A_a} shows the modulo of the dimensionless m.f. accelerations for different values of $\delta$. As expected,the depletion of the density merely results in weakening of the m.f. acceleration of stars in the core.

 is plotted.

\subsection{The effect of truncated pair-wise potential on density profile and m.f. acceleration}\label{subsec:trunc_Aphi}
Since the truncated acceleration does not include the effect of the cut-off of the density due to the assumption for DF, one may employ the truncated m.f. potential whose pair-wise potential takes zero value on scales of $r_{12}<\triangle$ implying non-existence of stars in the core. This is simply the case that one employs typical BBGKY hierarchy \ref{Eq.BBGKY_orth} and the g-Landau kinetic equation, \eqref{Eq.Generalized_Landau}, but the pairwise acceleration must be modified as follows
\begin{align}
\bmath{a}_{12}=-\nabla_{1}\left(\phi_{12}\Theta(r_{12}-\triangle)\right)
\end{align}
This modification results in Poisson equation,\eqref{Eq.grained_Poisson_Phi}, for m.f. potential and one may employ the density for truncated potential 
\begin{align}
\bar{n}^{(\Phi)}(\bmath{r},t)=\frac{1}{4\pi}\int\text{d}\Omega' \left(1-\triangle\hat{r'}\cdot\nabla\right)n(\bmath{r}-\triangle\hat{r'})
\end{align}
As explained in Appendix \ref{Appendix:density_phi12}, the density $\bar{n}^{(\Phi)}$ for a (quasi-static) spherical symmetric system reduces to the following expected form
\begin{align}
\bar{n}^{(\Phi)}(r)=\frac{(r+\triangle)n(r+\triangle)+(r-\triangle)n(\mid r -\triangle\mid)}{2}\label{Eq.n_Phi}
\end{align}
where it is to be noted that the factor $(r-\triangle)$ takes its absolute value only as the argument of the raw density. 

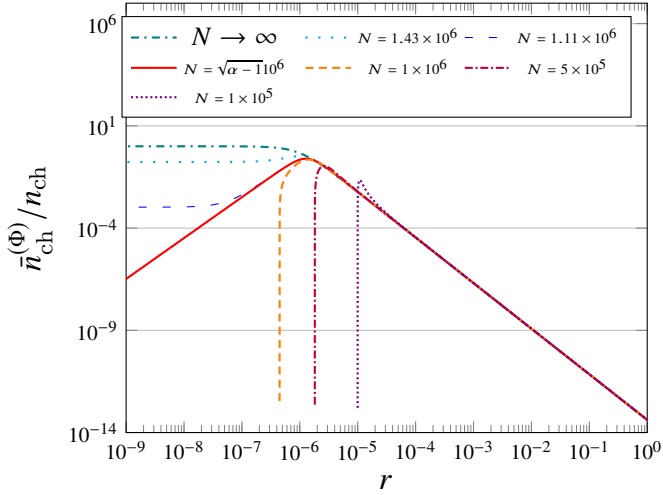
\begin{figure}
	\centering
	\begin{tikzpicture}[scale=1]
	\begin{loglogaxis}[ xlabel=\Large{$r$},ylabel=\Large{$\bar{n}^{(\Phi)}_\text{ch}/n_\text{ch}$},xmin=1e-9,xmax=1e0,ymin=1e-14,ymax=1e7, legend pos=north east,   ymajorgrids, legend columns=3, legend style={/tikz/column 2/.style={column sep=5pt,},},]
	\addplot [color = teal,mark=no,thick, dashdotted] table[x index=0, y index=1]{nphi_rho_1to4.txt}; 
	\addlegendentry{\large{$N\to\infty$}};
	\addplot [color = cyan, thick, loosely dotted] table [x index=0, y index=3]{nphi_rho_1to4.txt};
	\addlegendentry{$N =1.43\times10^{6}$};
	\addplot [color = blue, loosely dashed] table [x index=0, y index=4]{nphi_rho_1to4.txt};
	\addlegendentry{$N =1.11\times10^{6}$};
	\addplot [color = red, thick, solid] table [x index=0, y index=5]{nphi_rho_1to4.txt};
	\addlegendentry{$N = \sqrt{\alpha-1}10^{6}$};
	\addplot [color = orange, thick, densely dashed] table [x index=0, y index=1]{nphi_5.txt};
	\addlegendentry{$N =1\times10^{6}$};
	\addplot [color = purple, thick, densely dashdotted] table [x index=0, y index=1]{nphi_6.txt};
	\addlegendentry{$N =5\times10^{5}$};
	\addplot [color = violet, thick, densely dotted] table [x index=0, y index=1]{nphi_7.txt};
	\addlegendentry{$N =1\times10^{5}$};
	\end{loglogaxis}
	\end{tikzpicture}
	\caption{The density profile for truncated potential}
	\label{fig:density_Phi}
\end{figure}

For the modified Hubble model, the coarse-grained density is shown in Figure \ref{fig:density_Phi}. The density depletes in the core if the core size is close to or larger than the Landau radius ($\delta \leq \triangle$) in a similar way to the density for the truncated acceleration. The central density, however, reaches zero at the critical core radius
\begin{align}
r_\text{cr}=\triangle\sqrt{\alpha -1}\approx 0.9017\triangle
\end{align}
 which can be analytically determined by the condition
 \begin{align}
\bar{n}^{(\Phi)}_\text{ch}(\triangle)+\triangle\frac{\partial \bar{n}^{(\Phi)}_\text{ch}(\triangle)}{\partial r}(\triangle)=0
 \end{align}
If the core size smaller than the critical core size ($\delta>r_\text{cr}$), the position of maximum density becomes closer to the distance of $\triangle$ from the center and the size of zero-density region becomes larger. 

\begin{table}
	\caption{Zeros of the modified Hubble density profile due to the effect of truncation on the pair-wise potential.}
	\label{tabel:zeros}
	\begin{tabular}{|l|l|c|}
		\hline
		$\delta $                      & total number of stars            & the location of zero \\ 
		\hline
		$(\alpha-1)^{-1/2}$     &  $(\alpha-1)^{1/2}\times 10^{6}$         & 0                \\
		$1$       &  $1\times10^{6}$          &      $4.4677960\times10^{-7} $              \\
		$2$       &  $5\times10^{5}$           &  $1.8141649\times10^{-6} $                \\
		$10$       &  $1\times10^{5}$            &   $0.99749102\times10^{-5}    $          \\
		$50$        &  $2\times10^{4}$            & $4.9996532\times10^{-5}  $              \\
		$100$        &    $1\times10^{4}$           & $0.99998521\times10^{-4}    $               \\
		$1000$       &   $1\times10^{3}$           & $0.99999991\times10^{-3}   $               \\
		\hline
	\end{tabular}\\
\end{table}

Lastly

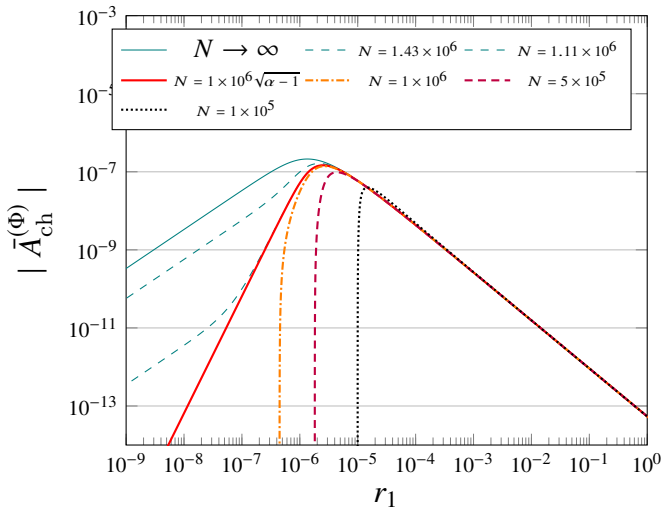
\begin{figure}
	\centering
	\begin{tikzpicture}[scale=1]
	\begin{loglogaxis}[ xlabel=\Large{$r_{1}$},ylabel=\Large{$\mid \bar{A}^{(\Phi)}_\text{ch}\mid$},xmin=1e-9,xmax=1,ymin=1e-14,ymax=1e-3, legend pos=north east,   ymajorgrids, yminorgrids, minor grid style={orange,dashed},legend columns=3 ]
	\addplot [color =teal,mark=circle,thin, solid] table[x index=0, y index=1]{Aphio.txt}; 
	\addlegendentry{\large{$N\to\infty$}};
	\addplot [color =teal,mark=circle,thin, dashed] table[x index=0, y index=1]{Aphi07.txt}; 
	\addlegendentry{$N=1.43\times 10^{6}$};
       \addplot [color =teal,mark=circle,thin, dashed] table[x index=0, y index=1]{Aphi09.txt}; 
	\addlegendentry{$N=1.11\times 10^{6}$};
      \addplot [color =red,mark=no,thick, solid] table[x index=0, y index=1]{Aphi_al.txt}; 
	\addlegendentry{$N=1\times 10^{6}\sqrt{\alpha-1}$};
      \addplot [color =orange,mark=no,thick, densely dashdotted] table[x index=0, y index=1]{Aphi1.txt}; 
	\addlegendentry{$N=1\times 10^{6}$};
      \addplot [color =purple,mark=no,thick, densely dashed] table[x index=0, y index=1]{Aphi2.txt}; 
	\addlegendentry{$N=5\times 10^{5}$};
      \addplot [color =black,mark=no,thick, densely dotted] table[x index=0, y index=1]{Aphi10.txt}; 
	\addlegendentry{$N=1\times 10^{5}$};
	\end{loglogaxis}
	\end{tikzpicture}
	\caption{mean field acceleration of stars due to the effect of the truncation on pair-wise potential}
	\label{fig:A_phi}
\end{figure}

\section{Discussion}

\subsection{first order approximation of the exact form}
\begin{figure}
	\centering
	\begin{tikzpicture}[scale=1]
	\begin{loglogaxis}[ xlabel=\Large{$r$},ylabel=\Large{$\bar{n}^{(comp)}_\text{ch}/n_\text{ch}$},xmin=1e-9,xmax=1e0,ymin=1e-14,ymax=1e7, legend pos=north east,   ymajorgrids, legend columns=3, legend style={/tikz/column 2/.style={column sep=5pt,},},]
	\addplot [color = teal,mark=no,thick, dashdotted] table[x index=0, y index=1]{Ncompo.txt}; 
	\addlegendentry{\large{$N\to\infty$}};
	\addplot [color = cyan, thick, loosely dotted] table [x index=0, y index=1]{Ncomp07.txt};
	\addlegendentry{$N =1.43\times10^{6}$};
	\addplot [color = blue, loosely dashed] table [x index=0, y index=1]{Ncomp09.txt};
	\addlegendentry{$N =1.11\times10^{6}$};
	\addplot [color = red, thick, solid] table [x index=0, y index=1]{Ncomp_al.txt};
	\addlegendentry{$N = \sqrt{\alpha-1}10^{6}$};
	\addplot [color = orange, thick, densely dashed] table [x index=0, y index=1]{Ncomp1.txt};
	\addlegendentry{$N =1\times10^{6}$};
	\addplot [color = purple, thick, densely dashdotted] table [x index=0, y index=1]{Ncomp2.txt};
	\addlegendentry{$N =5\times10^{5}$};
	\addplot [color = violet, thick, densely dotted] table [x index=0, y index=1]{Ncomp10.txt};
	\addlegendentry{$N =1\times10^{5}$};
	\end{loglogaxis}
	\end{tikzpicture}
	\caption{The density profile for the first order approximation of the exact form}
	\label{fig:density_1st_o}
\end{figure}
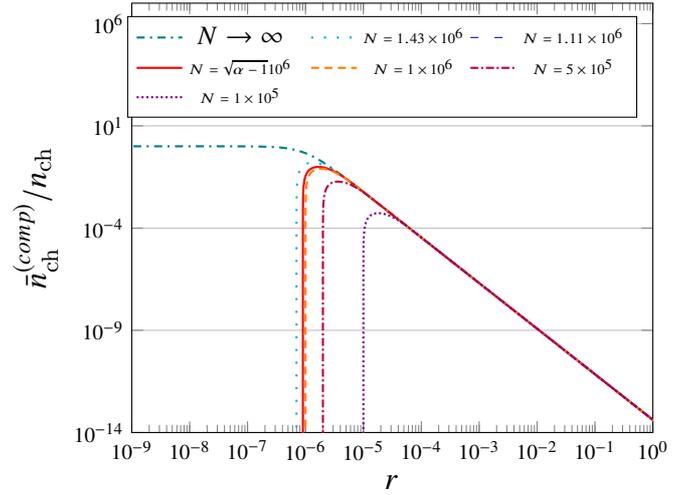 

\subsection{comparison}
The difference between the two description between two g-Landau kinetic equations lies on the way two stars interact within the Landau sphere; one is characterised by zero pair-wise acceleration and another by zero pairwise potential. The former is an approximated description for the latter. The advantage and disadvantage are summarized in Table

\begin{table}
	\caption{The decrease rate of the Coulomb logarithm, $100(1-B/\ln[N])$, due to the effect of discreteness.}
	\label{tabel:comparison}
	\begin{tabular}{|l|l|}
		\hline
		               &  zero acceleration in the core                      \\ 
		               & dispersion approximation ($\varv_{12}=<\varv>$)\\
		\hline
\cellcolor{yellow}		advantage      & Total number dependence              \\
\cellcolor{yellow}		               & fixed B.C for numerical integration  \\
\cellcolor{yellow}		                &                      \\
 \hline
\cellcolor{cyan}		disadvantage    & approximated form    \\
\cellcolor{cyan}	                   & coarse-grained density is integral form    \\
\cellcolor{cyan}	                   & weak density depletion in the core    \\
		\hline
		\hline
		              &  zero potential in the core                        \\ 
		               & dispersion approximation ($\varv_{12}=<\varv>$)\\
\hline
\cellcolor{yellow}advantage      & Total number dependence                 \\
 \cellcolor{yellow}              &  exact formulation                   \\
 \cellcolor{yellow}              &                    no integral in coarse-grained density                  \\
               \hline
\cellcolor{cyan}disadvantage    & definition of total number                          \\
\cellcolor{cyan}                &Changeable B.C. for numerical integration          \\
\cellcolor{cyan}                &                \\
		\hline
\hline
               &  typical FP (Landau)    \\ 
               & cold approximation ($\varv_{12}=0$)\\
\hline
\cellcolor{yellow} advantage      & well-known                \\
\hline
\cellcolor{cyan}disadvantage    &  inconsistent (collision term takes dispersion approximation)     \\
\hline
		
	\end{tabular}\\
\end{table}

\subsection{The condition to hold the assumption}
The failure of the m.f. acceleration $\bmath{A}^{a}$ originate from the definition for the total number $\int f(1,t)\text{d}t=N$.

yet, one may take the partial derivative over the equation with respect to $\bmath{r}_{1}$, which results in 'Poisson equation' for total number
\begin{align}
\nabla_{1}^{2}N^{*}(\bmath{r}_{1},t)=\frac{4\pi\triangle}{3}(n^{(a)}(\bmath{r}_{1},t)-n^{(\Phi)}(\bmath{r}_{1}))
\end{align}
Hence, the 'source' of total number appears if the number densities $n^(a)(\bmath{r}_{1},t)$ and $n^(\Phi)(\bmath{r}_{1},t)$ do not coincide with each other. Of course, it is the case when one considers there exist stars inside the Landau sphere for truncated pair-wise acceleration $\bmath{a}_{12}^{\triangle}$.

\subsection{Comparison between two acceleration}
As shown by the modified Hubble model, one can obtain the expected result by use of truncated DF; the core density is depleted meaning the stars can be found less likely in the Landau sphere centered at the origin. Typical numerical integration of FP equation takes two steps (i) Fokker-Planck step and (ii) Poisson step. Solving FP equation for the slow relaxation process can be return only fine-grained density (DF) while solving Poisson equation can return only the m.f. potential (acceleration) determined by the coarse-grained density. The depletion of the core density occurs unless the input density from the FP equation has a singular solution whose power is stronger than negative of one.

As discussed in section \ref{sec:truncated_BBGKY}, the truncated m.f. acceleration is correct only for density up to order of $N\bar{n}$ though, it showed a depletion effect due to the truncation of the domain integral i.e. no acceleration on scales smaller than the Landau sphere. On one hand, one needs to discuss the difference between the truncated- acceleration and potentials. The truncated acceleration is an outcome based on the assumption that two stars can not approach each other closer than the Landau radius while the truncated m.f. potential is an outcome from different assumption that stars can approach limitlessly in configuration space but pairwise potential has lower cut-off on scale of the Landau radius. The both of acceleration and potential showed expected depletion of density on scales smaller than the Landau while the truncated potential prohibit developments of density; the truncated potential may be suitable to avoid an infinite density problem. 

\subsection{For numerical integration}
\section{conclusion}\label{sec:conclusion}
Basic scenario for evolution of dense star clusters allows not to consider the effect of discreteness (finite $N$ effect). Yet, fundamental free parameter is only $N$ for ideal star clusters. In the present paper, it is clearly shown that even star clusters at self-similar epoch can be affected by the total number of stars though truncation of m.f. acceleration by assuming that star can not approach closer than the scale of Landau distance.

In section \ref{sec:truncated_BBGKY}, the \emph{weakly-coupled} DF and truncated DF were introduced to model the evolution of star clusters and the corresponding BBGKY hierarchies were derived. The lower limit of 'discreteness' fluctuations in m.f. acceleration of stars was cut-off on scale of the Landau distance.  It was especially shown that the truncated DF could hold the conservation of total- number and energy of stars if one employs the weakly-coupled DF while the use of weakly-coupled DF means that one neglects the effect of strong encounters and keeps losing a few of stars in evolution of the star in relaxation evolution of star clusters.

In section \ref{sec:complete_WC}, beginning with the BBGKY hierarchy for the weakly-coupled DF (assuming no stars can approach each other closer than the Landau radius), the g-Landau equation with 'completely'-weak-coupling approximation was derived. The mathematical formulation based on the weakly-coupled DF is corresponding to a kinetic formulation of the classical works \citep{Chandra_1943a,Takase_1950} and gives a correct treatment of m.f. potential for the cut-off problem and estimatation of the loss of the stars into the Landau sphere. 

In section \ref{sec:grainess_effect}, employing the simple relation between the system size and Landau radius, equation \eqref{Eq.simple_R_ro}, the effect of truncated phase-space volume elements in the g-Landau collision integral term weakens typical Coulomb logarithm, $\ln[N]$, for relatively small-number star cluster ($N=10^{5}$) by 14.0 $\%$ and for relatively large-number clusters ($N=10^{7}$) by 10.0 $\%$. Another effect of discreteness appears in the Poisson equation where the truncated volume elements simply corresponds with a coarse-graining of the density of stars by isotropising the density at the Landau radius. 

In later papers, the following generalization and application will be done. The 'actual' Landau distance is essentially naive to the relative speed of test star to field one, hence the distance must be correctly handled without velocity dispersion approximation. This necessitate even reapplying \citep{Grad_1958}'s method to velocity space, following the basic result of \citep{Takase_1950}. The method should be also extended to a case in which the system includes the effect of the strong encounter that can be described by the surface-integral terms in equation \eqref{Eq.BBGKY_truncated} neglected in the present work.
 
\section*{Acknowledgements}
I appreciate my adviser Carlo Lancellotti for allowing me to pursue this topic.




\bibliographystyle{mnras}
\bibliography{science}

\begin{thebibliography}{}
\makeatletter
\relax
\def\mn@urlcharsother{\let\do\@makeother \do\$\do\&\do\#\do\^\do\_\do\%\do\~}
\def\mn@doi{\begingroup\mn@urlcharsother \@ifnextchar [ {\mn@doi@}
  {\mn@doi@[]}}
\def\mn@doi@[#1]#2{\def\@tempa{#1}\ifx\@tempa\@empty \href
  {http://dx.doi.org/#2} {doi:#2}\else \href {http://dx.doi.org/#2} {#1}\fi
  \endgroup}
\def\mn@eprint#1#2{\mn@eprint@#1:#2::\@nil}
\def\mn@eprint@arXiv#1{\href {http://arxiv.org/abs/#1} {{\tt arXiv:#1}}}
\def\mn@eprint@dblp#1{\href {http://dblp.uni-trier.de/rec/bibtex/#1.xml}
  {dblp:#1}}
\def\mn@eprint@#1:#2:#3:#4\@nil{\def\@tempa {#1}\def\@tempb {#2}\def\@tempc
  {#3}\ifx \@tempc \@empty \let \@tempc \@tempb \let \@tempb \@tempa \fi \ifx
  \@tempb \@empty \def\@tempb {arXiv}\fi \@ifundefined
  {mn@eprint@\@tempb}{\@tempb:\@tempc}{\expandafter \expandafter \csname
  mn@eprint@\@tempb\endcsname \expandafter{\@tempc}}}

\bibitem[\protect\citeauthoryear{Aarseth \& Heggie}{Aarseth \&
  Heggie}{1998}]{Aarseth_1998}
Aarseth S.~J.,  Heggie D.~C.,  1998, \mn@doi [Monthly Notices of the Royal
  Astronomical Society] {10.1046/j.1365-8711.1998.01521.x}, 297, 794

\bibitem[\protect\citeauthoryear{{Adkins}}{{Adkins}}{2013}]{Adkins_2013}
{Adkins} G.~S.,  2013, preprint, \href
  {http://adsabs.harvard.edu/abs/2013arXiv1302.1830A} {} (\mn@eprint {arXiv}
  {1302.1830})

\bibitem[\protect\citeauthoryear{Ambartsumian}{Ambartsumian}{1938}]{Ambartsumian_1938}
Ambartsumian V.,  1938, in translated in eds J Goodman and P Hut Dynamics of
  Star Clusters, Proc. IAU Symp. pp 521--524

\bibitem[\protect\citeauthoryear{Balescu}{Balescu}{1997}]{Balescu_1997}
Balescu R.,  1997, Statistical Dynamics: matter out of equilibrium.
Imperial College press, \mn@doi{10.1142/p036}, \url
  {https://doi.org/10.1142%2Fp036}

\bibitem[\protect\citeauthoryear{Binney \& Tremaine}{Binney \&
  Tremaine}{2011}]{Binney_2011}
Binney J.,  Tremaine S.,  2011, Galactic Dynamics.
Princeton university press

\bibitem[\protect\citeauthoryear{Bittencourt}{Bittencourt}{2004}]{Bittencourt_2004}
Bittencourt J.~A.,  2004, Fundamentals of Plasma Physics.
Springer New York, \mn@doi{10.1007/978-1-4757-4030-1}, \url
  {https://doi.org/10.1007%2F978-1-4757-4030-1}

\bibitem[\protect\citeauthoryear{Bose \& Janaki}{Bose \&
  Janaki}{2012}]{Bose_2012}
Bose A.,  Janaki M.,  2012, \mn@doi [The European Physical Journal B]
  {10.1140/epjb/e2012-30357-x}, 85

\bibitem[\protect\citeauthoryear{Cercignani}{Cercignani}{1972}]{Cercignani_1972}
Cercignani C.,  1972, \mn@doi [Transport Theory and Statistical Physics]
  {10.1080/00411457208232538}, 2, 211

\bibitem[\protect\citeauthoryear{Cercignani}{Cercignani}{1988}]{Cercignani_1988}
Cercignani C.,  1988, The Boltzmann Equation and Its Applications.
Springer New York, \mn@doi{10.1007/978-1-4612-1039-9}, \url
  {https://doi.org/10.1007%2F978-1-4612-1039-9}

\bibitem[\protect\citeauthoryear{Cercignani}{Cercignani}{2008}]{Cercignani_2008}
Cercignani C.,  2008, in , Boltzmann's Legacy.
European Mathematical Society, pp 107--127, \mn@doi{10.4171/057-1/8}, \url
  {https://doi.org/10.4171/057-1/8}

\bibitem[\protect\citeauthoryear{Chandrasekhar}{Chandrasekhar}{1942}]{Chandra_1942}
Chandrasekhar S.,  1942, Principles of Stellar Dynamics.
 Vol. 1, The University of Chicago Press

\bibitem[\protect\citeauthoryear{{Chandrasekhar}}{{Chandrasekhar}}{1943}]{Chandra_1943a}
{Chandrasekhar} S.,  1943, \mn@doi [\apj] {10.1086/144517}, \href
  {http://adsabs.harvard.edu/abs/1943ApJ....97..255C} {97, 255}

\bibitem[\protect\citeauthoryear{Chang}{Chang}{1992}]{Chang_1992}
Chang Y.,  1992, \mn@doi [Physics of Fluids B: Plasma Physics]
  {10.1063/1.860279}, 4, 313

\bibitem[\protect\citeauthoryear{Chavanis}{Chavanis}{2012}]{Chavanis_2012}
Chavanis P.-H.,  2012, \mn@doi [Physica A: Statistical Mechanics and its
  Applications] {10.1016/j.physa.2012.02.019}, 391, 3680

\bibitem[\protect\citeauthoryear{Chavanis}{Chavanis}{2013}]{Chavanis_2013a}
Chavanis P.-H.,  2013, \mn@doi [Astronomy {\&} Astrophysics]
  {10.1051/0004-6361/201220607}, 556, A93

\bibitem[\protect\citeauthoryear{Cohen, Spitzer  \& Routly}{Cohen
  et~al.}{1950}]{Cohen_1950}
Cohen R.~S.,  Spitzer L.,   Routly P.~M.,  1950, \mn@doi [Phys. Rev.]
  {10.1103/PhysRev.80.230}, 80, 230

\bibitem[\protect\citeauthoryear{Cohn}{Cohn}{1979}]{Cohn_1979}
Cohn H.,  1979, \mn@doi [{Astrophysical Journal}] {10.1086/157587}, 234, 1036

\bibitem[\protect\citeauthoryear{Gilbert}{Gilbert}{1968}]{Gilbert_1968}
Gilbert I.~H.,  1968, \mn@doi [The Astrophysical Journal] {10.1086/149616},
  152, 1043

\bibitem[\protect\citeauthoryear{{Gilbert}}{{Gilbert}}{1971}]{Gilbert_1971}
{Gilbert} I.~H.,  1971, \mn@doi [\apss] {10.1007/BF00649188}, \href
  {http://adsabs.harvard.edu/abs/1971Ap%26SS..14....3G} {14, 3}

\bibitem[\protect\citeauthoryear{Goodman}{Goodman}{1983}]{Goodman_1983}
Goodman J.,  1983, \mn@doi [The Astrophysical Journal] {10.1086/161161}, 270,
  700

\bibitem[\protect\citeauthoryear{Goodman}{Goodman}{1984}]{Goodman_1984}
Goodman J.,  1984, \mn@doi [The Astrophysical Journal] {10.1086/161859}, 278,
  893

\bibitem[\protect\citeauthoryear{Grad}{Grad}{1958}]{Grad_1958}
Grad H.,  1958, in , Handbuch der Physik / Encyclopedia of Physics.
Springer Berlin Heidelberg, pp 205--294, \mn@doi{10.1007/978-3-642-45892-7_3},
  \url {https://doi.org/10.1007%2F978-3-642-45892-7_3}

\bibitem[\protect\citeauthoryear{Green}{Green}{1956}]{Green_1956}
Green M.~S.,  1956, \mn@doi [The Journal of Chemical Physics]
  {10.1063/1.1743132}, 25, 836

\bibitem[\protect\citeauthoryear{Griffel}{Griffel}{2002}]{Griffel_2002}
Griffel D.~H.,  2002, Applied functional analysis.
Courier Corporation

\bibitem[\protect\citeauthoryear{Heggie \& Hut}{Heggie \&
  Hut}{2003}]{Heggie_2003}
Heggie D.,  Hut P.,  2003, The Gravitational Million-Body Problem.
Cambridge University Press ({CUP}), \mn@doi{10.1017/cbo9781139164535}, \url
  {http://dx.doi.org/10.1017/cbo9781139164535}

\bibitem[\protect\citeauthoryear{{H{\'e}non}}{{H{\'e}non}}{1961}]{Henon_1961}
{H{\'e}non} M.,  1961, Annales d'Astrophysique, \href
  {http://adsabs.harvard.edu/abs/1961AnAp...24..369H} {24, 369}

\bibitem[\protect\citeauthoryear{Ipser \& Semenzato}{Ipser \&
  Semenzato}{1983}]{Ipser_1983}
Ipser J.~R.,  Semenzato R.,  1983, \mn@doi [The Astrophysical Journal]
  {10.1086/161196}, 271, 294

\bibitem[\protect\citeauthoryear{Jeans}{Jeans}{1902}]{Jeans_1902}
Jeans J.~H.,  1902, \mn@doi [Philosophical Transactions of the Royal Society A:
  Mathematical, Physical and Engineering Sciences] {10.1098/rsta.1902.0012},
  199, 1

\bibitem[\protect\citeauthoryear{Kandrup}{Kandrup}{1980}]{Kandrup_1980}
Kandrup H.~E.,  1980, \mn@doi [Physics Reports] {10.1016/0370-1573(80)90015-0},
  63, 1

\bibitem[\protect\citeauthoryear{Kandrup}{Kandrup}{1981a}]{Kandrup_1981}
Kandrup H.,  1981a, \mn@doi [The Astrophysical Journal] {10.1086/158709}, 244,
  316

\bibitem[\protect\citeauthoryear{Kandrup}{Kandrup}{1981b}]{Kandrup_1981_a}
Kandrup H.~E.,  1981b, \mn@doi [The Astrophysical Journal] {10.1086/158775},
  244, 1039

\bibitem[\protect\citeauthoryear{Kandrup}{Kandrup}{1986}]{Kandrup_1986}
Kandrup H.~E.,  1986, \mn@doi [Astrophysics and Space Science]
  {10.1007/bf00656047}, 124, 359

\bibitem[\protect\citeauthoryear{Kandrup}{Kandrup}{1988}]{Kandrup_1988}
Kandrup H.~E.,  1988, \mn@doi [Monthly Notices of the Royal Astronomical
  Society] {10.1093/mnras/235.4.1157}, 235, 1157

\bibitem[\protect\citeauthoryear{Kaufman}{Kaufman}{1960}]{Kaufman_1960}
Kaufman A.~N.,  1960, in Witt C.~D.,  Detoeuf J.,  eds, , La th\'{e}orie des
  gaz neutres et ionis\'{e}s.
John Wiley {\&} Sons, Inc., pp 293--353

\bibitem[\protect\citeauthoryear{King}{King}{1966}]{King_1966}
King I.~R.,  1966, \mn@doi [The Astronomical Journal] {10.1086/109857}, 71, 64

\bibitem[\protect\citeauthoryear{King}{King}{1985}]{King_1985}
King I.~R.,  1985, in , Dynamics of Star Clusters.
Springer Netherlands, pp 1--17, \mn@doi{10.1007/978-94-009-5335-2_1}, \url
  {https://doi.org/10.1007%2F978-94-009-5335-2_1}

\bibitem[\protect\citeauthoryear{Lanford}{Lanford}{1981}]{Lanford_1981}
Lanford O.~E.,  1981, \mn@doi [Physica A: Statistical Mechanics and its
  Applications] {10.1016/0378-4371(81)90207-7}, 106, 70

\bibitem[\protect\citeauthoryear{Liboff}{Liboff}{1965}]{Liboff_1965}
Liboff R.~L.,  1965, \mn@doi [Physics of Fluids] {10.1063/1.1761390}, 8, 1236

\bibitem[\protect\citeauthoryear{Liboff}{Liboff}{1966}]{Liboff_1966}
Liboff R.~L.,  1966, \mn@doi [Physics of Fluids] {10.1063/1.1761692}, 9, 419

\bibitem[\protect\citeauthoryear{Liboff}{Liboff}{2003}]{Liboff_2003}
Liboff R.~R.,  2003, Kinetic Theory: Classical, Quantum, and Relativistic
  Descriptions.
Springer-Verlag New York, \mn@doi{10.1007/b97467}

\bibitem[\protect\citeauthoryear{Lifshitz \& Pitaevskii}{Lifshitz \&
  Pitaevskii}{1981}]{Lifschitz_1981}
Lifshitz E.,  Pitaevskii L.,  1981, Physical kinetics (Course of theoretical
  physics, Oxford.
Pergamon Press

\bibitem[\protect\citeauthoryear{Lugger, Cohn  \& Grindlay}{Lugger
  et~al.}{1995}]{Lugger_1995}
Lugger P.~M.,  Cohn H.~N.,   Grindlay J.~E.,  1995, \mn@doi [The Astrophysical
  Journal] {10.1086/175164}, 439, 191

\bibitem[\protect\citeauthoryear{Mayer \& MG}{Mayer \& MG}{1940}]{Mayer_1940}
Mayer J.,  MG M.,  1940, Statistical mechanics.
John Wiley {\&} Sons, Inc.

\bibitem[\protect\citeauthoryear{McQuarrie}{McQuarrie}{2000}]{McQuarrie_2000}
McQuarrie D.~A.,  2000, Statistical mechanics.
University Science Books

\bibitem[\protect\citeauthoryear{Mehrem}{Mehrem}{2011}]{Mehrem_2011}
Mehrem R.,  2011, \mn@doi [Applied Mathematics and Computation]
  {10.1016/j.amc.2010.12.004}, 217, 5360

\bibitem[\protect\citeauthoryear{{Merritt}}{{Merritt}}{2013}]{Merritt_2013}
{Merritt} D.,  2013, Dynamics and Evolution of Galactic Nuclei.
Princeton University Press

\bibitem[\protect\citeauthoryear{Montgomery \& Tidman}{Montgomery \&
  Tidman}{1964}]{Montgomery_1964}
Montgomery D.~C.,  Tidman D.~A.,  1964, Plasma kinetic theory.
McGraw-Hill Advanced Physics Monograph Series, New York: McGraw-Hill, 1964

\bibitem[\protect\citeauthoryear{Ogorodnikov}{Ogorodnikov}{1965}]{Ogorodnikov_1965}
Ogorodnikov K.,  1965, Dynamics of stellar systems.
Oxford: Pergamon, 1965, edited by Beer, Arthur

\bibitem[\protect\citeauthoryear{Retterer}{Retterer}{1979}]{Retterer_1979}
Retterer J.~M.,  1979, \mn@doi [The Astronomical Journal] {10.1086/112432}, 84,
  370

\bibitem[\protect\citeauthoryear{Saslaw}{Saslaw}{1985}]{Saslaw_1985}
Saslaw W.~C.,  1985, Gravitational Physics of Stellar and Galactic Systems.
Cambridge University Press ({CUP}), \mn@doi{10.1017/cbo9780511564239}, \url
  {http://dx.doi.org/10.1017/cbo9780511564239}

\bibitem[\protect\citeauthoryear{Severne \& Haggerty}{Severne \&
  Haggerty}{1976}]{Severne_1976}
Severne G.,  Haggerty M.~J.,  1976, \mn@doi [Astrophysics and Space Science]
  {10.1007/bf00642666}, 45, 287

\bibitem[\protect\citeauthoryear{Shevelko \& Tawara}{Shevelko \&
  Tawara}{2012}]{Shevelko_2012}
Shevelko V.,  Tawara H.,  eds, 2012, Atomic Processes in Basic and Applied
  Physics.
Springer Berlin Heidelberg, \mn@doi{10.1007/978-3-642-25569-4}, \url
  {https://doi.org/10.1007%2F978-3-642-25569-4}

\bibitem[\protect\citeauthoryear{Shoub}{Shoub}{1992}]{Shoub_1992}
Shoub E.~C.,  1992, \mn@doi [The Astrophysical Journal] {10.1086/171231}, 389,
  558

\bibitem[\protect\citeauthoryear{Spitzer}{Spitzer}{1988}]{Spitzer_1988}
Spitzer L.~S.,  1988, Dynamical Evolution of Globular Clusters.
Walter de Gruyter {GmbH}, \mn@doi{10.1515/9781400858736}, \url
  {http://dx.doi.org/10.1515/9781400858736}

\bibitem[\protect\citeauthoryear{Takahashi}{Takahashi}{1995}]{Takahashi_1995}
Takahashi K.,  1995, Publications of the Astronomical Society of Japan, 47, 561

\bibitem[\protect\citeauthoryear{{Takase}}{{Takase}}{1950}]{Takase_1950}
{Takase} B.,  1950, \pasj, \href
  {http://adsabs.harvard.edu/abs/1950PASJ....2....1T} {2, 1}

\bibitem[\protect\citeauthoryear{Trigger, Ershkovich, van Heijst  \&
  Schram}{Trigger et~al.}{2004}]{Trigger_2004}
Trigger S.~A.,  Ershkovich A.~I.,  van Heijst G. J.~F.,   Schram P. P. J.~M.,
  2004, \mn@doi [Physical Review E] {10.1103/physreve.69.066403}, 69

\bibitem[\protect\citeauthoryear{Wen \& Avery}{Wen \& Avery}{1985}]{Wen_1985}
Wen Z.-Y.,  Avery J.,  1985, \mn@doi [Journal of Mathematical Physics]
  {10.1063/1.526621}, 26, 396

\bibitem[\protect\citeauthoryear{Zeidler, Hackbusch, Schwarz  \& Hunt}{Zeidler
  et~al.}{2004}]{Zeidler_2004}
Zeidler E.,  Hackbusch W.,  Schwarz H.~R.,   Hunt B.,  2004, Oxford users'
  guide to mathematics.
Oxford University Press

\makeatother
\end{thebibliography}

\appendix
\begin{appendices}
\renewcommand{\theequation}{\Alph{section}.\arabic{equation}}

\section{BBGKY hierarchy for distribution function and scalings of physical quantities in stellar dynamics}\label{sec:basic_kinetics}
In sections \ref{subsec_Liouville} and \ref{subsec:s-tuple} fundamental concepts of kinetic theory are reviewed and in section \ref{subsec:scalings}  a scaling of orders of the magnitudes (OoM) of physical quantities to describe a star cluster and encounters is explained. In section \ref{subsec:trajectory} the trajectory of test star in encounters is explained. In section \ref{subsec:Log_col} the logarithmic divergences in collision- and wave- kinetic theories are explained.
\subsection{The $N$-body Liouville equation}\label{subsec_Liouville}

Consider a star cluster of $N$-'point' stars of equal masses $m$ interacting each other purely via and $r_{ij}\left(=\mid \bmath{r}_{i}-\bmath{r}_{j}\mid\right)$ is the distance between star $i$ at position $\bmath{r}_{i}$ and star $j$ at $\bmath{r}_{j}$. The Hamiltonian for the motions of stars in the system reads
\begin{equation}
H=\sum_{i=1}^{N}\left(\frac{\bmath{p}^{2}_{i}}{2m}+m\sum_{j>i}^{N}\phi(r_{ij})\right),\label{Eq.H}
\end{equation}
where $\bmath{p}_{i}(=m\bmath{\varv}_{i})$ is the momentum of star $i$ moving at velocity $\bmath{\varv}_{i}$. Assume the corresponding $6N$ Hamiltonian equations can be alternatively written in form of the $N$-body Liouville equation
\begin{align}
\frac{\text{d}F_{N}}{\text{d}t}=\left(\partial_{t}+\sum_{i=1}^{N}\left[\bmath{\varv}_{i}\cdot\nabla_{i}+\bmath{a}_{i}\cdot \bmath{\partial}_{i}\right]\right)F_{N}(1,\cdots,N,t)=0,\label{Eq.Liouville}
\end{align}
where the symbols for the operators are abbreviated by $\partial_{t}=\frac{\partial}{\partial t}$, $\nabla_{i}=\frac{\partial}{\partial \bmath{r}_{i}}$, and $\bmath{\partial}_{i}=\frac{\partial}{\partial \bmath{\varv}_{i}}$. The acceleration $\bmath{a}_{i}$ of star $i$ due to the pair-wise 

 and the Hamiltonian equation \eqref{Eq.H} in phase space (obviously) holds the same symmetry, meaning stars $1$, $\cdots$, $N$ are assumed identical and indistinguishable respectively.

\subsection{The $s$-tuple distribution function and correlation function}\label{subsec:s-tuple}
A reduced DF of stars in a star cluster is, in general, introduced in form of $s$-body (joint-probability) DF
\begin{equation}
F_{s}(1\cdots s, t)=\int F_N(1,\cdots,N, t)\quad\text{d}_{s+1}\cdots\text{d}_{N},\label{Eq.SbodyJointDF}
\end{equation} 
or in form of 

The $s$-tuple DF simplifies the relation of macroscopic quantities with irreducible $s$-body dynamical functions. For example, the total energy of the system at time $t$ may read
\begin{subequations}
	\begin{align}
	E(t)&=\int\cdots\int\sum_{i=1}^{N}\left(\frac{\bmath{p}^{2}_{i}}{2m}+m\sum_{j>i}^{N}\phi(r_{ij})\right)F_{N}(1\cdots N,t)\text{d}_1\cdots\text{d}_N,\\
	&=N\int \frac{\bmath{p}_{1}^{2}}{2m}F_{1}(1,t)\text{d}_{1}+m\frac{N(N-1)}{2}\int\phi(r_{12})F_{2}(1,2,t)\text{d}_1\text{d}_2,\\
	&=\int \frac{\bmath{p}_{1}^{2}}{2m}f_{1}(1,t)\text{d}_{1}+\frac{m}{2}\int\phi(r_{12})f_{2}(1,2,t)\text{d}_1\text{d}_2.\label{totE}
	\end{align}
\end{subequations}
where the symmetry of permutation between two phase-space points for both the Hamiltonian and the $s$-body DF are applied. The total energy $E(t)$ can turn into a more physically meaningful form by introducing typical $s$-ary DFs to understand the effect of correlation between stars, as follows.
\begin{align*}
\quad	f(1,t) \quad&:\text{(unary) DF}  \\
\quad	f(1,2,t) \quad &:\text{binary DF} \\
\quad	g(1,2,t) \quad &:\text{(binary) correlation function} \\
\quad	f(1,2,3,t) \quad &:\text{ternary DF}\\
\quad	T(1,2,3,t) \quad &:\text{ternary correlation function}
\end{align*}
 As proved under the weak-coupling approximation by \cite{Liboff_1965,Liboff_1966} and employed by \cite{Gilbert_1968,Gilbert_1971}, the correlation function $g(i,j,t)$ has the anti$-$normalization property for self-gravitating systems
\begin{equation}
\int g(i,j,t)\text{d}_{i}=\int g(i,j,t)\text{d}_{j}=0. \qquad(i,j=1, 2, \text{ or } 3\quad \text{with } i\neq j).\label{Eq.Anti-norm}
\end{equation}
Employing the correlation function $g(1,2,t)$, the total energy, equation \eqref{totE}, may be rewritten as
\begin{align}
&E(t)=\int \frac{\bmath{p}_{1}^{2}}{2m}f_{1}(1,t)\text{d}_{1}+U_{\text{m.f.}}(t)+U_{\text{cor}}(t),\label{Eq.E}
\end{align}
where
\begin{subequations}
	\begin{align}
	&U_{\text{m.f.}}(t)=\frac{m}{2}\int\Phi(\bmath{r}_{1},t)f(1,t)\text{d}_{1},\label{Eq.U_id}\\
	&U_{\text{cor}}(t)=\frac{m}{2}\int\phi(r_{12})g(1,2,t)\text{d}_{1}\text{d}_{2},\label{Eq.U_cor}
	\end{align}
\end{subequations}
and the self-consistent gravitational m.f. potential is defined as
\begin{align}
&\Phi(\bmath{r}_{1},t)=\left(1-\frac{1}{N}\right)\int \phi(r_{12})f(2,t)\text{d}_{2},\label{Eq.Phi}
\end{align}
where the factor $\left(1-\frac{1}{N}\right)$ is also the effect of discreteness; the m.f. potential on a star is due to $(N-1)$-field stars \citep[e.g.][]{Kandrup_1986}. Also, the corresponding self-consistent gravitational m.f. acceleration of star 1 reads
\begin{align}
&\bmath{A}(\bmath{r}_{1},t)=-\left(1-\frac{1}{N}\right)\int \nabla_{1}\phi(r_{12})f(2,t)\text{d}_{2}.\label{Eq.A}
\end{align}

\subsection{Scaling of the order of magnitudes of physical quantities}\label{subsec:scalings}
Section \ref{subsec:basic_scaling} explains the basic scalings of physical quantities employed in the present work and in section \ref{subsection:scaling_strong} the scaling associated with strong two-body encounters.

\subsubsection{basic scalings}\label{subsec:basic_scaling}
One needs two scaling parameters for non-divergent kinetic theory; the discreteness parameter, $1/N$, and the distance $r_{12}$ between two stars (say, star 1 is test star at $\bmath{r}_{1}$ and star 2 is one of field stars at $\bmath{r}_{2}$.). The fundamental scaling of physical quantities associated with the discreteness parameter follows the scaling employed in \citep[][Appendix A]{Chavanis_2013a} except for the correlation function $g(1,2,t)$ (equation \eqref{Eq.scale_g(1,2,t)}). For $r_{12}$, following the scaling of the order of magnitudes (OoM) of physical quantities for classical electron-ion plasmas \citep[][pg 22]{Montgomery_1964}, one may classify the effective distance of two-body Newtonian interaction and m.f. acceleration into the following four ranges of distance between two stars depending on the magnitude of forces due to the accelerations on test star in a star cluster system;
\begin{enumerate}
	\item m.f.(many-body) interaction \qquad\quad \quad $d$$<r_{12}<$ $R$
	\item weak m.f.(many-body) interaction \quad $ a_\text{BG}$ $<r_{12}<$ $d$
	\item weak two-body interaction \qquad \quad \quad $r_\text{o}$ $<r_{12}<$ $a_\text{BG}$
	\item strong two-body interaction \qquad  \quad\quad$ 0<r_{12}<$ $r_\text{o}$
\end{enumerate}
where $R$ is the characteristic size of a finite star cluster (e.g. the Jeans length and tidal radius), $d$ the average distance of stars in the system, $r_\text{o}$ the 'conventional' Landau radius (to be explained in section \ref{subsection:scaling_strong}) and $a_\text{BG}$ the Boltzmann-Grad(BG) radius. The BG radius separates the distance $r_{12}$ at which two-body encounters are dominant from those at which the effect of m.f. acceleration (many-body encounters) is dominant; $a_\text{BG}$ corresponds with the scaling of Boltzmann-Grad limit \citep{Grad_1958}\footnote{It is to be noted that the BG radius is in essence the same as the '\emph{encounter radius} \citep{Ogorodnikov_1965}' to separate the encounter and passage of stars.}. For relaxation processes in plasmas \citep{Montgomery_1964}, the BG radius $a_\text{BG}$ is of no essence since the fundamental mathematical formulation assumes homogeneous plasmas and the Thermodynamic limit,
\begin{align}
n=N/V\to\mathcal{O}(1)\qquad (\text{with}\quad V \to \infty\quad \text{and}\quad N \to \infty),
\end{align} 
where $V$ is the system volume of plasmas.

In the present work, the 'Landau radius' $r_{90}$ is newly defined as the closest spatial separation of two stars when the impact parameter of test star is equal to the Landau distance $b_{90}$ (the impact parameter to deflect test star thorough an encounter by $90^\text{o}$ from the original direction of motion);
\begin{subequations}
	\begin{align}
	&r_{90}(\varv_{12}(-\infty))=\frac{b_{90}(\varv_{12}(-\infty))}{1+\sqrt{2}},\label{Eq.r_90}\\
	&b_{90}(\varv_{12}(-\infty))=\frac{2Gm}{\varv^{2}_{12}(-\infty)}\label{Eq.b_90},
	\end{align}
\end{subequations}
where $\varv_{12}(-\infty)$ is the relative speed between star 1 and star 2 before encounter.
Refer to Tables \ref{table:scale} and \ref{table:scale_fig} for the scalings of basic physical quantities and Appendix \ref{Appendix:interaction} for how some of the scalings, especially the ranges of distances, could be determined. The characteristic time scales of the relevant evolution of DFs and correlation function are defined as 
\begin{subequations}
	\begin{align}
	&\frac{1}{t_\text{dyn}}\simeq \left|\frac{\bmath{\varv}_{1}\cdot\nabla_{1} f(1,t)}{f(1,t)}\right|,\\
	&\frac{1}{t_\text{sec}}\simeq \left|\frac{1}{f(1,t)}\left(\frac{\partial f(1,t)}{\partial  t}\right)\right|,\\
	&\frac{1}{t_\text{cor}}\simeq \left|\frac{1}{g(1,2,t)}\left(\frac{\partial  g(1,2,t)}{\partial  t}\right)\right|.
	\end{align} 
\end{subequations}

\begin{table}\centering
	\caption{A scaling of the order of magnitudes of physical quantities associated with the evolution of a star cluster that has not gone through a core-collapse. The scaling will be especially employed for a completely weakly-coupled- and weakly-inhomogeneous- star clusters in sections \ref{sec:complete_WC} and \ref{sec.WC_strong} respectively, whose density contrast is much less than the order of $N$. The OoM are scaled by $N$ and $r_{12}$ except for the correlation time $t_\text{cor}$, which needs the change in velocity, $\delta\bmath{\varv}_{a}\left(=\int\bmath{a}_{12}\text{d}t_\text{cor}\right)$, due to Newtonian two-body interaction.}
	\begin{tabular}{l l}
		\hline
		quantities & order of magnitude\\
		\hline
		$t_\text{r}$ & $\sim N/\ln[N],$\\
		$f(1,t), t_\text{sec}$ & $\sim N,$\\
		$\bmath{A}_{1},R, m,\bmath{\varv}_{1},\varv_{12},\bmath{r}_{1},t_\text{dyn}$&$\sim 1,$\\
		$d$ &$\sim 1/N^{1/3},$\\
		$a_\text{BG}$ & $\sim1/N^{1/2}$\\
		$G, r_\text{o}, K_{n}$&$\sim 1/N,$\\
		\hline
		$g(1,2,t)$&$ \sim N/r_{12},\quad$ for $r_\text{o}<r_{12}<R$ \\
		&$\sim N^{2},\qquad$ for $r_{12}<r_\text{o}$\\
		$\bmath{a}_{12}$&$\sim 1/(r_{12}^{2}N) $\\
		$t_\text{cor}$, $\delta\bmath{\varv}_{A}\left(=\int\bmath{A}_{1}\text{d}t_\text{cor}\right)$&$\sim r_{12}$ $\quad$ for  $r_\text{o}<r_{12}<R$\\
		&$\sim\delta \bmath{\varv}_{12} r_{12}^{2}N$ $\quad$ for $r_{12}<r_\text{o}$\\
		\hline
	\end{tabular}\label{table:scale}
\end{table}

\begin{table}
	\caption{A scaling of physical quantities according to the effective interaction range of Newtonian interaction accelerations and close encounter}
	\begin{tikzpicture}
	\filldraw[thin, fill=pink!20!white, opacity=0.1, shading=ball] (0,1)--(8,1) arc(0:30:8);
	\filldraw[thin, fill=pink!30!white, opacity=0.2, shading=ball] (0,1)--(6,1) arc(0:30:6);
	\filldraw[thin, fill=pink!40!white, opacity=0.3, shading=ball] (0,1)--(4,1) arc(0:30:4);
	\filldraw[thin, fill=pink!50!white, opacity=0.4, shading=ball] (0,1)--(2,1) arc(0:30:2);
	\draw(6,1) arc(0:30:6);
	\draw(4,1) arc(0:30:4);
	\draw(2,1) arc(0:30:2);
	\draw[densely dashed, thick] (8,0.9)--(8,-0.4);
	\draw[densely dashed, thick] (6,0.9)--(6,-0.4);
	\draw[densely dashed, thick] (4,0.9)--(4,-0.4);
	\draw[densely dashed, thick] (2,0.9)--(2,-0.4);
	\node[below] at (8,-0.6) {$R$};
	\node[below] at (6,-0.6) {$d$};
	\node[below] at (4,-0.6) {$a_\text{BG}$};
	\node[below] at (2,-0.6) {$r_\text{o}$};
	\node at (0,-0.7) {${r}_{12}$};
	\node at (1,0.5) {strong 2-body};\node at (1,0.2) {(Boltzmann)};
	\node at (3,0.5) {weak 2-body};\node at (3,0.2) {(Landau)};
	\node at (5,0.5) {weak m.f};\node at (5,0.2) {(g-Landau)};
	\node at (7,0.5) {m.f.(many-body)};\node at (7,0.2) {(g-Landau)};
	
	\node[below] at (8,-1.3) {$\sim 1$};
	\node[below] at (6,-1.2) {$\sim 1$};
	\node[below] at (4,-1.2) {$\sim 1$};
	\node[below] at (2,-1.2) {$\sim 1$};
	\node[below] at (0,-1.3) {$\bmath{A}_{1}$};
	
	\node[below] at (8,-2.3) {$\sim N$};
	\node[below] at (6,-2.2) {$\sim N^{4/3}$};
	\node[below] at (4,-2.2) {$\sim N^{3/2}$};
	\node[below] at (2,-2.2) {$\sim N^{2}$};
	\node[below] at (0,-2.3) {$g(1,2,t)$};
	
	\node[below] at (8,-3.0) {$\sim 1/N$};
	\node[below] at (6,-2.9) {$\sim N^{-1/3}$};
	\node[below] at (4,-3) {$\sim 1$};
	\node[below] at (2,-3) {$\sim N $};
	\node[below] at (0,-3) {$\bmath{a}_{12}$};
	
	\node[below] at (8,-3.7) {$\sim 1/N$};
	\node[below] at (6,-3.6) {$\sim N^{-2/3}$};
	\node[below] at (4,-3.6) {$\sim N^{-1/2}$};
	\node[below] at (2,-3.6) {$\sim 1 $};
	\node[below] at (0,-3.6) {$\delta\varv_{\bmath{a}}$};
	
	\node[below] at (8,-4.4) {$\sim 1$};
	\node[below] at (6,-4.3) {$\sim N^{-1/3}$};
	\node[below] at (4,-4.3) {$\sim N^{-1/2}$};
	\node[below] at (2,-4.3) {$\sim N^{-1}$};
	\node[below] at (0,-4.3) {$\delta\varv_{\bmath{A}}$, $t_\text{cor}$};
	
	\shadedraw[inner color=orange, outer color=yellow, draw=black] (0,1) circle (0.1cm);
	\node [above] at (0,1.1) {star 1};
	\shadedraw[inner color=orange, outer color=yellow, draw=black] (6.2,1.2) circle (0.1cm);
	\node [right] at (6.3,1.4) {star 2};
	\end{tikzpicture}\label{table:scale_fig}
\end{table}
\subsubsection{Close encounter and encounters with large-deflection angle and large-speed change }\label{subsection:scaling_strong}
A special focus of the scaling is the Landau radius $r_\text{o}$, equation \eqref{Eq.r_90}, since it especially depends on the relative speed between two stars. A mathematically correct treatment on the Landau distance has been discussed for Newtonian interaction \citep{Retterer_1979,Ipser_1983,Shoub_1992} and Coulombian one \citep{Chang_1992}, until then one had simplified the Landau distance by approximating the relative speed $\varv_{12}$ to the velocity dispersion $<\varv>$ of the system; the 'conventional' Landau- distance, $b_\text{o}$, and and radius, $r_\text{o}$, are defined as
\begin{align}
&b_{90}\simeq\frac{2Gm}{<\varv>^{2}}\equiv b_\text{o},\label{Eq.thermo_approx_b}\\
&r_\text{o}\equiv\frac{b_\text{o}}{1+\sqrt{2}}.\label{Eq.thermo_approx_r}
\end{align}
Assume the dispersion speed may be determined by the Virial theorem for a finite spherical star cluster of radius of $R$ as follows
\begin{align}
<\varv>\equiv c\sqrt{\frac{GmN}{R}},\label{Eq.<v>}
\end{align}
where $c$ is a constant and the radius $R$ may be the Jeans length or tidal radius to hold the finiteness of the system size. Simple examples for the value of the constant $c$ are; $c=\sqrt{3/5}$ if the system is finite and spatially homogeneous and $c$ is order of unity if the system follows the King model \citep{King_1966}. In the present paper, the dispersion approximation is still employed since it simplifies the scaling of the Landau distance without losing the essential property of strong encounters. Employing equations \eqref{Eq.thermo_approx_b}, \eqref{Eq.thermo_approx_r} and \eqref{Eq.<v>},  one finds the relation between the system radius and the Landau distance as follows
\begin{align}
\frac{r_\text{o}}{R}=\frac{2}{1+\sqrt{2}}\frac{1}{c^{2}N}. \label{Eq.ratio_R_ro}
\end{align}

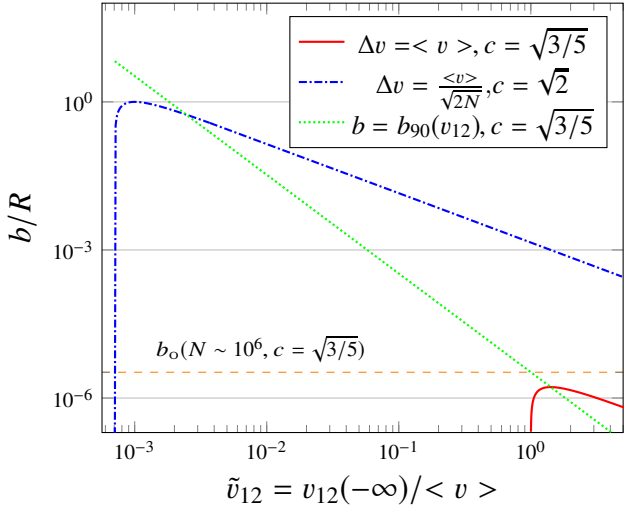
\begin{figure}
	\centering
	\begin{tikzpicture}[scale=1]
	\begin{loglogaxis}[ xlabel=\Large{$\tilde{\varv}_{12}=\varv_{12}(-\infty)/<\varv>$},ylabel=\Large{$b/R$},xmin=0.0008/sqrt(2),xmax=5,ymin=2e-7,ymax=100, legend pos=north east, ytick={1,1e-3,1e-6}, minor ytick={3.33e-6}, ymajorgrids, yminorgrids, minor grid style={orange,dashed} ]
	\addplot [color = red ,mark=no,thick, solid] table[x index=0, y index=1]{b90_velocity_dep_3.txt}; 
	\addlegendentry{\large{$\Delta\varv=<\varv>,c=\sqrt{3/5}$}};
	\addplot [color = blue ,mark=no,thick, densely dashdotted ] table[x index=0, y index=4]{b90_velocity_dep.txt};
	\addlegendentry{\large{$\Delta\varv=\frac{<\varv>}{\sqrt{2N}}$,$c=\sqrt{2}$}}; 
	\addplot [color = green ,mark=no,thick, densely dotted] table[x index=0, y index=5]{b90_velocity_dep.txt}; 
	\addlegendentry{\large{$b=b_{90}(\varv_{12}),c=\sqrt{3/5}$}};
	\addplot [color = blue ,mark=no,thick, densely dashdotted ] table[x index=0, y index=4]{b90_velocity_dep_2.txt};
	\addplot [color = green ,mark=no,thick, densely dotted ] table[x index=0, y index=5]{b90_velocity_dep_2.txt}; 
	\node at (9e-3,10e-6){$b_\text{o}(N\sim10^{6},c=\sqrt{3/5})$}; 
	\end{loglogaxis}
	\end{tikzpicture}
	\caption{The normalised-speed-dependence of the impact parameter $b$  for different changes of speed $\Delta \varv$ of test star, where $b(\tilde{\varv}_{12})=2R[\left({\varv}_{12}(-\infty)/\Delta\varv\right)^{2}-1]^{1/2}/[\sqrt{N}c\tilde{\varv}_{12}]^{2}$ (, which can be derived from equations \eqref{Eq_close_encounter2} and \eqref{Eq_close_encounter3} \emph{without} the dispersion approximation, or see equation 13 of \citep{Shoub_1992}). The solid curve ($\Delta\varv=<\varv>$) separates encounter (scattering) events; the encounter  events described by the region below the solid curve are strong encounters, while those above the curve is weak one. The dotted line separates large-angle- and small-angle deflection of star. In the present paper considering the two-body encounters are spatially-local events due to m.f. acceleration being dominant on larger scales than $a_\text{BG}$, the local weak- and strong- encounters are defined only on the region below the horizontal-grid line of $a_\text{BG}$. Especially, the encounters defined above the Landau distance $r_\text{o}$ are to be called \emph{distant} two-body encounter and encounters below $r_\text{o}$ are \emph{close} one. }
	\label{fig:Landau_distance}
\end{figure}

As discussed in \citep{Shoub_1992}, one may separate the impact parameter $b$ of encounter into weak- and strong- deflections following the change in speed of test star through an two-body encounter (Figure \ref{fig:Landau_distance}). In general, kinds of 'strong' two-body encounter is either of large-angle $\left(\gtrsim 90^\text{o}\right)$ deflection and large-speed $\left(\gtrsim <\varv>\right)$ change of test star. In figure \ref{fig:Landau_distance}, the former is described by the region below the dotted curve and the latter is described by the region below the solid curve.  For mathematical convenience, \citep{Shoub_1992} chose the speed change of $ <\varv>/\sqrt{2N}$ (the dashdotted curve on figure \ref{fig:Landau_distance}) to delimit the strong- and weak- encounters at which the change in speed of test star is the same order of the speed change caused by a distant field star on the system-size scale via Newtonian pair-wise acceleration. (Of course one does not have to delimit the encounters since even weak deflections can be described by the Boltzmann-collision description.). However, in more realistic systems, the upper limit of impact parameter for two-body encounter is approximately the BG radius, $a_\text{BG}$, up to which the Boltzmann-collision (collision kinetic) description may be defined. Also actual strong encounters occur only on scales smaller than $b_\text{o}$ (at most $R/(c^{2}N)$)and the slowest relative speed $\varv_{12}(-\infty)$ that causes a large change in speed is equal to the speed dispersion $<\varv>$. Correspondingly, the maximum impact parameter that includes both of strong encounter and large-angle-deflection encounter is the conventional Landau distance, equation \eqref{Eq.thermo_approx_b}. Hence, one may scale the maximum impact parameter as the conventional Landau distance;
\begin{align}
b_\text{max}= b_\text{90}\approx b_\text{o}\sim\mathcal{O}(1/N).\label{Eq.thermo_approx_2}
\end{align}
Equation \eqref{Eq.thermo_approx_2} can be reasonable under the following condition. If one neglects the contribution from energetic stars faster than the escape speed of the system ($\approx 2 <\varv>$), the strong-encounter is 'localized' around the velocity dispersion in relative-speed spaces. Only in this sense, one may employ a dispersion approximation for the relative speed
\begin{align}
\varv_{12}\approx <\varv>\sim\mathcal{O}(1)\label{Eq.thermo_approx_3}
\end{align}

Following the discussion above, one may understand that choosing the conventional Landau distance $b_\text{o}$ as the maximum impact parameter of encounters and employing the dispersion approximation are to focus on each \emph{close-strong} encounter that includes the effects of large-angle-deflection- and strong- encounters on scales smaller than the distance $b_\text{o}$.\footnote{Technically speaking, the relative speed $\tilde{\varv}_{12}$ is a function of the speed $\Delta \varv$ and impact parameter $b$ in finding the explicit form of Boltzmann equation as done in \citep{Shoub_1992} hence the domain of $\tilde{\varv}_{12}$ for strong encounter is to be determined by the relation between the change in speed $\Delta \varv$ and the dispersion $<\varv>$, while one does not need to resort to the serious discussion for the scaling purpose.}

Since the relative-speed dependence of the impact parameter may be loosely neglected, one can define the 'Knudsen number' for each close strong encounter by
\begin{align}
K_{n}= R\tilde{n} \sigma^{(p)},\label{Eq.Kn}
\end{align}
where the momentum-transfer cross section \citep[e.g.][]{McQuarrie_2000,Bittencourt_2004,Shevelko_2012}, $\sigma^{(p)}$, due to the close-strong encounters and the corresponding Newtonian-'scattering' relation may be characterised respectively by
\begin{subequations}
	\begin{align}
	&\sigma^{(p)}\approx 2\pi\int^{b_\text{o}}_{0} \left(1-\cos\theta\right)b\text{d}b, \label{Eq_close_encounter1}\\
	&\tan\frac{\theta}{2}\approx\frac{b_\text{o}}{b},\label{Eq_close_encounter2}\\
	&\Delta\varv_{1}\approx<\varv>\sin\frac{\theta}{2},\label{Eq_close_encounter3}
	\end{align}
\end{subequations}
where  $\theta$ is the deflection angle of the unperturbed trajectory of a test star due to each close strong encounter and  the impact parameter $b$ reaches the conventional Landau distance $b_\text{o}$ at $\theta=\pi/2$. It is to be noted that $b_\text{o}$ and $b$ in equations \eqref{Eq_close_encounter1}, \eqref{Eq_close_encounter2} and \eqref{Eq_close_encounter3} do not explicitly or even implicitly depend on relative speed $\varv_{12}$ since the equations are a direct consequence of equations \eqref{Eq.thermo_approx_2} and \eqref{Eq.thermo_approx_3} (i.e. the dispersion approximation).  Also, the mean density $\tilde{n}$ in equation \eqref{Eq.Kn} may be still the order of $N$ since the ejection- and evaporation- rates may be longer than the time scale of secular evolution are less significant except for the core-collapse stage \citep[e.g][]{Spitzer_1988,Binney_2011}. The order of the cross section $\sigma^{(p)}$ is
\begin{align}
&\sigma^{(p)}=2\ln[2]\pi b_\text{o}^{2}\sim\mathcal{O}(1/N^{2}),
\end{align}
correspondingly
\begin{align}
K_{n}= Rn \sigma^{(p)}\sim\mathcal{O}(1/N).
\end{align}
Hence, one may consider the close-strong two-body encounter is also characterized by the discreteness parameter, $1/N$. The Knudsen number $K_{n}$ may be understood as approximately the possibility of finding test star experiencing a close-strong encounter in the 'Landau sphere' (the sphere of radius $r_\text{o}$ around a field star) on dynamical-time scale, $t_\text{dyn}$. In more actual situation, the Landau sphere, of course, does not correctly isolate strong encounters from weak ones; weak encounters may occur even in the Landau sphere due to the relative-speed dependence of the impact parameter. Exactly speaking, the Landau sphere must be exploited to separate close two-body encounters from distant ones, or collision kinetic description (two-body encounters) from wave one (many-body encounters). The latter helps one to understand the importance of the truncated m.f. acceleration $\bmath{A}^{\blacktriangle}(\bmath{r}_{1},t)$ due to the insignificance of the m.f. acceleration on small spatial scales in the secular evolution of a finite system;
\begin{align}
\bmath{A}^{\blacktriangledown}(\bmath{r}_{1},t)&=-\left(1-\frac{1}{N}\right)\int_{{r}_{12}<r_\text{o}} \nabla_{1}\phi(r_{12})f(2,t)\text{d}_{2},\label{Eq.A_down}\\
&\sim \mathcal{O}(1/N).\nonumber
\end{align}
where the the scaling $\partial_{t}f(1,t)\sim\mathcal{O}(1)$ is to be recalled. 

\subsection{Trajectories of a test star}\label{subsec:trajectory}

The complete (Lagrangian) trajectory of star $i$ can be discussed by taking the sum of the m.f. acceleration of star $i$ due to smooth m.f. potential force and the Newtonian pair-wise acceleration via interaction with star $j$;
\begin{subequations}
	\begin{align}
	&\bmath{r}_{i}(t)=\bmath{r}_{i}(t-\tau)+\int_{t-\tau}^{t}\bmath{\varv}_{i}\left(t'\right)\text{d}t',\hspace{15pt} (i\neq j =1,2)\\
	&\bmath{\varv}_{i}(t)=\bmath{\varv}_{i}(t-\tau)+\int_{t-\tau}^{t}\left[\bmath{a}_{ij}\left(t'\right)+\bmath{A}_{i}\left(t'\right)\right]\text{d}t'.
	\end{align}\label{Eq.characteristics_general}
\end{subequations}
One can approximate the complete trajectory to a simpler form in each range of distance between stars $i$ and $j$, following the scaling of section \ref{subsec:scalings}. At distances $r_{ij}<a_\text{BG}$ where two-body Newtonian interaction dominates the other effects, the trajectory perfectly follows a pure Newtonian two-body problem
\begin{subequations}
	\begin{align}
	&\bmath{r}_{i}(t)=\bmath{r}_{i}(t-\tau)+\int_{t-\tau}^{t}\bmath{\varv}_{i}\left(t'\right)\text{d}t',\\
	&\bmath{\varv}_{i}(t)=\bmath{\varv}_{i}(t-\tau)+\int_{t-\tau}^{t}\bmath{a}_{ij}\left(t'\right)\text{d}t'.
	\end{align}\label{Eq.characteristics_two_body}
\end{subequations}
At relatively short distances $(r_{ij}\lesssim r_\text{o})$, the trajectory due to a strong-close encounter may be considered as local Newtonian interaction between two stars (i.e. the Boltzmann two-body collision description if one includes the Markovian approximation)
\begin{subequations}
	\begin{align}
	&\bmath{r}_{12}(t)=\bmath{r}_{12}(t-\tau)+\int_{t-\tau}^{t}\bmath{\varv}_{12}\left(t'\right)\text{d}t',\\
	&\bmath{R}=\frac{\bmath{r}_{1}+\bmath{r}_{2}}{2}\approx \bmath{r}_{1},\\
	&\bmath{r}_{1}=\bmath{r}_{1}(t-\tau)+\int_{t-\tau}^{t}\frac{\bmath{\varv}_{1}\left(t'\right)+\bmath{\varv}_{2}\left(t'\right)}{2}\text{d}t',\\
	&\bmath{\varv}_{i}(t)=\bmath{\varv}_{i}(t-\tau)+\int_{t-\tau}^{t}\bmath{a}_{ij}\left(t'\right)\text{d}t'.
	\end{align}\label{Eq.characteristics_v_jump}
\end{subequations}
At intermediate distances $(r_\text{o}<< r_{ij}\lesssim a_\text{BG})$, the trajectory due to two-body weak-distant encounters may take rectilinear motion local in space with weak-coupling limit
\begin{subequations}
	\begin{align}
	&\bmath{r}_{12}(t)=\bmath{r}_{12}(t-\tau)+\bmath{\varv}_{12}\tau,\\
	&\bmath{R}\approx \bmath{r}_{1},\\
	&\bmath{\varv}_{i}(t)=\bmath{\varv}_{i}(t-\tau).
	\end{align}\label{Eq.characteristics_rectilinear}
\end{subequations}
Lastly at long distances $(a_\text{BG}<<r_{ij}<R)$, the trajectory due to many-body weak-distant encounter may purely follows the motion of star under the effect of m.f. acceleration with weak-coupling limit
\begin{subequations}
	\begin{align}
	&\bmath{r}_{i}(t)=\bmath{r}_{i}(t-\tau)+\int_{t-\tau}^{t}\bmath{\varv}_{i}\left(t'\right)\text{d}t',\\
	&\bmath{\varv}_{i}(t)=\bmath{\varv}_{i}(t-\tau)+\int_{t-\tau}^{t}\bmath{A}_{i}\left(t'\right)\text{d}t'.
	\end{align}\label{Eq.characteristics_m.f.}
\end{subequations}

\section{An explanation for the order of magnitude of the effective distance of Newtonian interaction potentials}\label{Appendix:interaction}
In the Appendix, the scaling of the OoM of the effective interaction range of accelerations of stars due to Newtonian potentials are explained following the ranges below;
\begin{enumerate}
	\item m.f. (many-body) interaction \qquad  $a_\text{BG}$$<r_{12}<$ $R$
	\item weak two-body interaction \quad  $ r_\text{o}$ $<r_{12}<$ $a_\text{BG}$
\end{enumerate}
where the average distance of stars is neglected since it is not of essence in the present work. To find the discussion for the Landau radius $r_\text{o}$, refer to section \ref{subsection:scaling_strong}.
 
\subsection{The threshold between (i) and (ii)}
The transition between ranges (i) and (ii) is the radius of encounter \citep{Ogorodnikov_1965}, at which the order of the irregular force is compatible with that of m.f. potential force. In range (i), star 1 can be accelerated by the total of Newtonian interaction forces due to the rest of stars as follows
\begin{align}
\bmath{a}_{1}=-\sum_{k=2}^{N}\frac{Gm}{r_{1k}^{3}} (\bmath{r}_{1}-\bmath{r}_{k}).
\end{align}
As assumed in \citep{Kandrup_1981} and Appendix \ref{sec.many_to_two}, the acceleration due to many-body encounters (with $(N-1)$-stars) may be roughly replaced by the acceleration due to the smooth self-consistent m.f. acceleration of star 1;
\begin{align}
\bmath{A}_{1}=-\left(1-\frac{1}{N}\right)\int \frac{Gm}{r_{12}^{3}}(\bmath{r}_{1}-\bmath{r}_{2})f(2,t)\text{d}_{2}.
\end{align}
Some stars, however, can occasionally enter range (ii), then the main cause of acceleration of star 1 is due to the pair-wise Newtonian potential, equation \eqref{Eq.phi}, from star 2
\begin{align}
\bmath{a}_{12}=-\frac{Gm}{r_{12}^{3}} (\bmath{r}_{1}-\bmath{r}_{2}).
\end{align}

Employing the scaling $G\sim\mathcal{O}(1/N)$ for fixed finite stellar masses $m\sim\mathcal{O}(1)$ and fixed momenta $\bmath{p}_{1}\sim\bmath{p}_{2}\sim\mathcal{O}(1)$ as explained in section \ref{subsec:scalings}, the two accelerations are scaled as
\begin{subequations}
	\begin{align}
	&\bmath{a}_{12}\sim \frac{1}{N}\frac{1}{r_{12}^{2}},\label{Eq.scale_a_12}\\
	&\bmath{a}_{1}\sim R\sim\mathcal{O}(1).\label{Eq.scale_A_1}
	\end{align}
\end{subequations}
By equating the two accelerations, equations \eqref{Eq.scale_a_12} and \eqref{Eq.scale_A_1}, the threshold between m.f. (many-body) and two-body interaction forces is obtained
\begin{align}
&r_{12}\sim \frac{1}{N^{1/2}}\sim a_\text{BG}.
\end{align}

\subsection{The size of a cluster in (i)}
Assume the size of a star cluster corresponds with the Jean length. The celebrated Jeans instability \citep{Jeans_1902} of a self-gravitating system may be discussed even at kinetic-equation level for collisionless \citep[e.g.][]{Binney_2011} and collisional \citep[e.g.][]{Trigger_2004} self-gravitating systems assuming the dynamical stability condition as follows
\begin{align}
\bmath{\varv}_{1}\cdot\nabla_{1}+\bmath{A}_{1}\cdot\bmath{\partial}_{1}\sim \frac{\bmath{\varv}_{1}}{R}- \frac{Gmnr_{12}}{\varv_1}=0.
\end{align}
where $R$ is the size of the stellar system and $n$ the average density of the system. Due to the unscreened gravitational potential, the interaction range $r_{12}$ or the wavelength of fluctuation in m.f. potential can reach the system radius $R$ and may bring the system into an unstable state. The Jeans length occurs when the distance $r_{12}$ is compatible with the system radius $R$
\begin{align}
R\sim \sqrt{\frac{\varv^{2}_{1}}{Gmn}}\sim\mathcal{O}(1),
\end{align}
where scalings $G\sim\mathcal{O}(1/N)$ and $n\sim\mathcal{O}(N)$ are taken for fixed stellar mass $m\sim\mathcal{O}(1)$ and dispersion $<\varv>\sim\mathcal{O}(1)$.

\section{Derivation of BBGKY hierarchy for truncated distribution function}\label{Appendix_BBGHY}
In \citep{Cercignani_1988}, the derivation of the BBGKY hierarchy for the hard-sphere DFs was made in a mathematically strict manner, by employing the Gauss's lemma and integration-by-parts, while counting the correct patterns of combinations for the Gauss's lemma is confusing and the BBGKY hierarchy for the truncated DF is not shown explicitly. In the present section, the latter hierarchy is derived by exploiting integration-by-parts and a general Heaviside function
\begin{align}
\Theta(r_{ij}-\triangle)&=
\begin{cases}
1            & \text{if}\quad r_{ij}\geq \triangle,\\
0             & \text{otherwise}. \label{Eq.Heaviside}
\end{cases}\\
&\equiv\theta_{(i,j)},
\end{align}
together with the following mathematical identity
\begin{align}
\nabla_{i}\theta_{(i,j)}=\frac{\bmath{r}_{ij}}{r_{ij}}\delta(r_{ij}-\triangle).\label{Eq:delta_func}
\end{align}
Use of the Heaviside function $\Theta(r_{ij}-\triangle)$ may admit of violating a mathematical strictness in distribution theory; the product of two genralised functions may not be well-defined in the sense of distribution \citep[e.g.][]{Griffel_2002}, since the $N$-body distribution function $F_{N}(1,\cdots,t)$\citep{Cercignani_1988} and the function $\Theta(r_{ij}-\triangle)$ are both generalised functions, while one will find its convenience of exploiting the Heaviside function to derive the \citep{Cercignani_1972}'s hierarchy below. 

First define the following term
\begin{align}
I_{s}&\equiv\int_{\Omega_{s+1,N}}\text{d}_{s+1}\cdots\text{d}_{N} S(1,\cdots, N, t),\label{Eq.Is_def}
\end{align}
where $S(1,\cdots, N, t)$ is any function of arguments $\{1,\cdots, N, t\}$. Following the domain, equation \eqref{Int_Omega}, of integration for the truncated DF, one may explicitly express the term as follows
\begin{align}
I_{s}&=\int\text{d}_{s+1} \hspace{2pt} \theta_{(s+1,1)}\hspace{3pt}\cdots  \theta_{(s+1,s)}\nonumber\\
     &\times\int\text{d}_{s+2} \hspace{2pt}  \theta_{(s+2,1)}\hspace{5pt}\cdots  \theta_{(s+2,s)}\hspace{2pt}\theta_{(s+2,s+1)}\nonumber\\
&\hspace{60pt}\vdots\hspace{40pt}\vdots\hspace{30pt}\vdots\hspace{22pt}\ddots\nonumber\\  
     &\times\int\text{d}_{N-1}\hspace{2pt}  \theta_{(N-1,1)}\cdots  \theta_{(N-1,s)}\theta_{(N-1,s+1)} \cdots\theta_{(N-1,N-2)}\nonumber\\
     &\times\int\text{d}_{N} \hspace{13pt}  \theta_{(N,1)}\hspace{2pt}\cdots \theta_{(N,s)}\hspace{7pt}\theta_{(N,s+1)}\hspace{5pt} \cdots\hspace{5pt}\theta_{(N,N-2)} \theta_{(N,N-1)}\nonumber\\
&\times S(1,\cdots, N, t).\label{Eq.Is}
\end{align}
 
\subsection{Truncated integral over the terms $\sum_{i=1}^{N}\bmath{\varv}_{i}\cdot\nabla_{i}F_{N}(1,\cdots,N,t)$}
For the function $S(1,\cdots, N, t)=\sum_{i=1}^{N}\bmath{\varv}_{i}\cdot\nabla_{i}F_{N}(1,\cdots,N,t)$, the pattern of subscripts of the distance $r_{ij}$ in equation \eqref{Eq.Is} is simple; the number $1\leq i\leq s$ appears only as the first letter in subscript. Hence one may separate the summation in the function $S(1,\cdots,N,t)$ into case 1: $1\leq i\leq s$ and case 2: $s+1\leq i\leq N$.

\subsubsection{Case 1: $1\leq i\leq s$}
The goal of the present Appendix is to reduce the term $I_\text{s}$ associated with the terms $\bmath{\varv}_{i}\cdot\nabla_{i}F_{s}^{\triangle}(1,\cdots,s,t)$ by repeating integral-by-parts method. For the numbers $1\leq i\leq s$, define the following term
\begin{align}
I_{s}^{(1:s)}&\equiv\sum_{i=1}^{s}\int_{\Omega_{s+1,N}}\text{d}_{s+1}\cdots\text{d}_{N} \bmath{\varv}_{i}\cdot\nabla_{i}F_{N}(1,\cdots N,t).
\end{align}
Employing equation \eqref{Eq:delta_func}, one obtains
\begin{align}
&I_{s}^{(1:s)}\nonumber\\
&=\sum_{i=1}^{s}\bmath{\varv}_{i}\cdot\nabla_{i}F_{s}^{\triangle}(1,\cdots,s,t) \nonumber\\
&\quad-\sum_{i=1}^{s}\bmath{\varv}_{i}\cdot\sum_{j=s+1}^{N}\int\text{d}_{N}\int\text{d}_{N-1}\cdots\int\text{d}_{j}\cdots\int\text{d}_{s+2}\int\text{d}_{s+1} \nonumber\\
&\quad\times \theta_{(s+1,1)}\cdots  \theta_{(s+1,i)}\cdots \theta_{(s+1,s)}\nonumber\\
&\quad\times \theta_{(s+2,1)}\cdots  \theta_{(s+2,i)}\cdots \theta_{(s+2,s)}\hspace{2pt} \theta_{(s+2,s+1)}\nonumber\\
&\hspace{80pt}\vdots\hspace{65pt}\ddots\nonumber\\  
&\quad\times\theta_{(j,1)}\hspace{5pt}\cdots \hspace{5pt}\frac{\bmath{r}_{ij}}{r_{ij}}\delta_{(j,i)}\hspace{4pt}\cdots\hspace{2pt}\theta_{(j,s)} \hspace{20pt} \cdots  \hspace{10pt}\theta_{(j,j-1)}\nonumber\\
&\hspace{80pt}\vdots \hspace{100pt}\ddots\nonumber\\  
&\quad\times \theta_{(N-1,1)}\cdots  \theta_{(N-1,i)} \cdots\theta_{(N-1, s)}\hspace{10pt}\cdots\hspace{15pt}\theta_{(N-1,N-2)}\nonumber\\
&\quad\times \theta_{(N,1)}\hspace{5pt}\cdots \hspace{5pt} \theta_{(N,i)}\hspace{5pt} \cdots\theta_{(N, s)}\hspace{20pt}\cdots\hspace{40pt}\theta_{(N,N-1)}\nonumber\\
&\quad\times F_{N}(1,\cdots, N, t),\label{Eq.Is_nab_i}
\end{align}
where  $\delta_{(j,i)}\equiv\delta(r_{ij}-\triangle)$. Due to the delta function $\delta_{(j,i)}$, one can convert the volume integral into the surface integral
\begin{align}
&\int\text{d}_{j} \theta_{(j,1)}\cdots  \frac{\bmath{r}_{ij}}{r_{ij}}\delta_{(j,i)}\cdots\theta_{(j,j-1)}=\int \text{d}^{3}\bmath{p}_{j} \oint \text{d}\bmath{\sigma}_{ij}, \label{eq.delta1}
\end{align}
where the $\text{d}\bmath{\sigma}_{ij}$ is the surface element of a sphere of radius $\triangle$ with a radial unit vector $\frac{\bmath{r}_{ij}}{r_{ij}}$ around the position $\bmath{r}_{j}$. Employing equation \eqref{eq.delta1}, one obtains
\begin{align}
I_{s}^{(1:s)}&=\sum_{i=1}^{s}\left(\bmath{\varv}_{i}\cdot\nabla_{i}F_{s}^{\triangle}-\sum_{j=s+1}^{N}\int \text{d}^{3}\bmath{p}_{j} \oint \bmath{\varv}_{i}\cdot\text{d}\bmath{\sigma}_{ij}F^{\triangle}_{s+1}(1,\cdots, s+1, t)\right).\label{Eq.Is_nab_ii}
\end{align}

\subsubsection{Case 2: $s+1\leq i\leq N$}
Define the term $I_\text{s}$associated with the numbers $s+1\leq i\leq N$;
\begin{align}
I_{s}^{(s+1:N)}&\equiv\sum_{i=s+1}^{N}\int_{\Omega_{s+1,N}}\text{d}_{s+1}\cdots\text{d}_{N} \bmath{\varv}_{i}\cdot\nabla_{i}F_{N}(1,\cdots N,t).
\end{align}
To reduce the term $I_{s}^{(s+1:N)}$, one must modify equation \eqref{Eq.Is_nab_i} as follows 
\begin{align}
&I_{s}^{(s+1:N)}\nonumber\\
&=-\sum_{i=s+1}^{N}\sum_{j=1}^{s}\int \text{d}^{3}\bmath{p}_{i} \oint \bmath{\varv}_{i}\cdot\text{d}\bmath{\sigma}_{ij}F^{\triangle}_{s+1}(1,\cdots, s+1, t)\nonumber\\
&\quad+\sum_{i=s+1}^{N}\int\bmath{\varv}_{i}\cdot\nabla_{i}\left(\prod_{i=s+1}^{N}\prod_{j=1}^{i-1}\theta_{(j,i)} F_{N}(1,\cdots N,t)\right) \text{d}_{s+1}\cdots\text{d}_{N}\nonumber\\
&\quad-\sum_{i=s+1}^{N}\bmath{\varv}_{i}\cdot\sum_{j=s+1}^{N}\nonumber\\
&\quad\times\int\text{d}_{N}\int\text{d}_{N-1}\cdots\int\text{d}_{i}\cdots\int\text{d}_{j}\cdots\int\text{d}_{s+2}\int\text{d}_{s+1}  \nonumber\\
&\quad\times \theta_{(s+1,1)}\cdots \theta_{(s+1,s)}\nonumber\\
&\quad\times \theta_{(s+2,1)}\cdots  \theta_{(s+2,s)}\theta_{(s+2,s+1)}\nonumber\\
&\hspace{20pt}\vdots\hspace{45pt}\vdots\hspace{25pt}\vdots\hspace{20pt}\ddots\nonumber\\ 
&\quad\times\theta_{(i+1,1)}\cdots\theta_{(i+1,s)}\theta_{(i+1,s+1)} \cdots \theta_{(i+1,i)}\nonumber\\
&\hspace{20pt}\vdots\hspace{45pt}\vdots\hspace{25pt}\vdots\hspace{40pt}\vdots\hspace{20pt}\ddots\nonumber\\ 
&\quad\times\theta_{(j,1)}\cdots\hspace{5pt}\theta_{(j,s)}\hspace{5pt}\theta_{(j,s+1)} \cdots \hspace{5pt}\frac{\bmath{r}_{ij}}{r_{ij}}\delta_{(j,i)} \cdots \theta_{(j,j-1)}\nonumber\\
&\hspace{20pt}\vdots\hspace{45pt}\vdots\hspace{25pt}\vdots\hspace{40pt}\vdots\hspace{40pt}\vdots\hspace{15pt}\ddots\nonumber\\ 
&\quad\times \theta_{(N,1)} \cdots\theta_{(N, s)}\theta_{(N,s+1)} \cdots \theta_{(N,i)}  \cdots \theta_{(N,j-1)} \cdots\theta_{(N,N-1)}\nonumber\\
&\quad\times F_{N}(1,\cdots, N, t).\label{Eq.Is_nab_j}
\end{align}
where the first summation of terms is obtained in the same way as done for equation \eqref{Eq.Is_nab_ii}, but this time the functions of the displacement vector $\bmath{r}_{j}$ (associated with the latter subscript $j$ in the distance $r_{ij}$) was differentiated. The second summation of terms on the R.H.S. in equation \eqref{Eq.Is_nab_j} vanishes if one assumes the function  $F_{i}(1,\cdots,i,t)$ approaches rapidly enough to zero at the surface of the integrals. Since the delta function in the third summation of terms links two volume integrals to a surface integral, one obtains
\begin{align}
I_{s}^{(s+1:N)}&=\sum_{i=1}^{s}\sum_{j=s+1}^{N}\int \text{d}^{3}\bmath{p}_{j} \oint \bmath{\varv}_{j}\cdot\text{d}\bmath{\sigma}_{ij}F^{\triangle}_{s+1}(1,\cdots, s+1, t)\nonumber\\
&\quad+\sum_{i=s+1}^{N}\sum_{j=s+1}^{N}\int \text{d}_{i}\int \text{d}^{3}\bmath{p}_{j} \oint \bmath{\varv}_{j}\cdot\text{d}\bmath{\sigma}_{ij}\nonumber\\
&\qquad\qquad\qquad\times F^{\triangle}_{s+2}(1,\cdots, s+2, t),\label{Eq.I_s_s+1}
\end{align}
where the following relation is employed
\begin{align}
&\int\text{d}_{i}\int \text{d}_{j} \hspace{5pt}\theta_{(j,1)}\hspace{5pt}\cdots \hspace{5pt}\frac{\bmath{r}_{ij}}{r_{ij}}\delta_{(j,i)}\hspace{4pt}\cdots  \hspace{10pt}\theta_{(j,j-1)}\nonumber\\
&\quad=\int \text{d}_{i}\int \text{d}^{3}\bmath{p}_{j} \oint \bmath{\varv}_{j}\cdot\text{d}\bmath{\sigma}_{ij}.
\end{align}
Combining the results above, equation \eqref{Eq.I_s_s+1}, with the result of case 1 ($1\leq i \leq s$) and considering the dummy integral variables, one obtains
\begin{align}
I_{s}=&\sum_{i=1}^{s}\bmath{\varv}_{i}\cdot\nabla_{i}F_{s}^{\triangle}+\sum_{i=1}^{s}(N-s)\left[\int\text{d}^{3}\varv_{s+1}\oiint F_{s+1}^{\triangle} \bmath{\varv}_{i,s+1}\cdot\text{d}\bmath{\sigma}_{i,s+1}\right]\nonumber\\
&+\frac{(N-s)(N-s-1)}{2}\int\text{d}^{3}\varv_{s+2}\int\text{d}_{s+1}\nonumber\\
&\quad\times\oiint F_{s+2}^{\triangle}  \bmath{\varv}_{s+1,s+2}\cdot\text{d}\bmath{\sigma}_{s+1,s+2},
\end{align} 
where $\bmath{\varv}_{ij}=\bmath{\varv}_{i}-\bmath{\varv}_{j}$. Only the configuration space in the truncated DF must be deprived, hence the rest of treatment for the other terms in the Liouville equation is the same as for the standard BBGKY hierarchy \citep[e.g.][]{Lifschitz_1981,Saslaw_1985,McQuarrie_2000,Liboff_2003}, which results in equation \eqref{Eq.BBGKY_truncated} in terms of $s$-tuple DFs.

\section{Derivation of angle-averaged density profile for spherically symmetric system}\label{Appendix:density}
In Appendix \ref{Appendix:density}, the `one-center' density profile $n(\mid \bmath{r}_{1}-\triangle \hat{r}\mid)$ for spherically symmetric system is rewiritten as a functional of  $n(r_{1})$. Then, by use of special funcitons, the derivaiotns of the formulas $\bar{n}^{(a)}(r_{1})$, equation \eqref{Eq.nbar_formula}, and equation \eqref{Eq.n_Phi}  are shown in Appendix

\subsection{From $n(\mid \bmath{r}_{1}-\triangle \hat{r}\mid)$ to $n(r_{1})$}\label{Appendix:density}
One may rewrite the density profile $n(\mid \bmath{r}_{1}-\triangle \hat{r}\mid)$ in term of the density profile $n(r_{1})$. To do so, one can follow the method discussed in for hyperspherical harmonics \citep{Wen_1985} and it applies to three dimensional case using Guggenbaur polynomials (The present paper  slightly different defintion to directly employ spherical Bessel polynomials.). 
 
One would like to find the following form of density profile 
\begin{align}
n(\mid \bmath{r}-\bmath{r}'\mid)=\int\text{d}r''\int\text{d}\Omega'' r''^{2}n(r'')\delta (\bmath{r}''-\bmath{r}'+\bmath{r})
\end{align}
where the delta function in coordinate spaces, by use of inverse Fourier transformation, reads
\begin{align}
\delta(\bmath{r})=\frac{1}{8\pi^{3}}\int \text{d}^{3}\bmath{k}\text{e}^{i\bmath{k}\cdot\bmath{r}}\label{Eq.delta_func}
\end{align}
In case of three dimensional configuration spaces, one can find the plane wave $\text{e}^{i\bmath{k}\cdot\bmath{r}}$ in terms of the $l$-th order spherical Bessel function $j_{l}(kr)$ of first kind and the $l$-th order Legendre polynomials $P_{l}\left(\hat{r}\cdot\hat{k}\right)$
\begin{align}
\text{e}^{i\bmath{k}\cdot\bmath{r}}=\sum_{l=0}^{\infty}i^{l}(2l+1)j_{l}(kr)P_{l}\left(\hat{r}\cdot\hat{k}\right)\label{Eq.plane}
\end{align}
where $\hat{k}$ and $\hat{r}$ are unit vectors of  wavenumber vector $\bmath{k}$ and position $\bmath{r}$ of star 1.
In addition, one may further employ the addition theorem to expand the plane wave in temrs of the associated Legendre polynomials 
\begin{align}
P_{l}\left(\hat{r}\cdot\hat{k}\right)=\sum_{m=-l}^{l}\text{e}^{im(\phi_{r}-\phi_{k})}P^{m}_{l}\left(\cos\theta_{k}\right)P^{m}_{l}\left(\cos\theta_{r}\right)\frac{(l-m)!}{(l+m)!}\label{Eq.add_theo}
\end{align}
where  $\phi_{r}$ and $\phi_{k}$ are the azimuthal angles and $\theta_{r}$ and $\theta_{k}$ the polar angles of the vectors $\bmath{r}$ and $\bmath{k}$ respectively. By employing equations \eqref{Eq.delta_func}, \eqref{Eq.plane} and \eqref{Eq.add_theo} suceseeiviely, the density profile reduces to the following form
\begin{align}
n(\mid \bmath{r}-\bmath{r}'\mid)=&\sum_{l=0}^{\infty}\sum_{l'=0}^{\infty}\int\text{d}r''\int\text{d}\bmath{k} r''^{2}n(r'')i^{l-l'}(2l+1)\nonumber\\
&\times(2l'+1) j_{l}(kr)j_{l'}(kr')j_{0}(kr'')P_{l}\left(\hat{r}\cdot\hat{k}\right)P_{l'}\left(\hat{r'}\cdot\hat{k}\right)\label{Eq.n2}
\end{align}
where the following identity is employed
\begin{align}
\int\text{d}\Omega''_{r}P_{l}\left(\hat{r''}\cdot\hat{k}\right)=4\pi\delta_{l,0}. \label{Eq.ave_Legendre}
\end{align}
where $\Omega''_{r}$ is the solid angle of the vector $\bmath{r}''$. To simplify equation \eqref{Eq.n2}, after employing the addition theorem again, use the following formula
\begin{align}
\int\text{d}\Omega_{k}P_{l}\left(\hat{r}\cdot\hat{k}\right)P_{l}\left(\hat{r'}\cdot\hat{k}\right)=\frac{4\pi}{2l+1}\delta_{l,l'}P_{l}\left(\hat{r'}\cdot\hat{r}\right). 
\end{align}
where$\Omega_{k}$ is the solid angle of the vector $\bmath{}$. Then, one obtains the following formula
\begin{align}
&n(\mid \bmath{r}-\bmath{r}'\mid)=\sum^{\infty}_{l=0}a_{l}(r,r')P_{l}\left(\hat{r'}\cdot\hat{r}\right)\\
&a_{l}(r,r')\equiv\frac{2(2l+1)}{\pi}\int\text{d}r'' r''^{2}n(r'')J_{0ll}(r,r',r'')\\
&J_{0ll}(r,r',r'')\equiv\int\text{d}k k^{2} j_{l}(kr)j_{l}(kr')j_{0}(kr'')
\end{align}

\subsection{From  $n^{(a)}(r)$ to $\bar{n}(r)$}
Employing equation \eqref{Eq.ave_Legendre},  the coarse-grained density profile $\bar{n}^{(a)}(r)$ can read
\begin{align}
&\bar{n}^{(a)}(\mid \bmath{r}-\bmath{r}'\mid)=a_{0}(r,r')P_{l}=\frac{2}{\pi}\int^{\infty}_{0}r''^{2}n(r'')J_{000}(r,r',r'')\\
&J_{000}(r,r',r'')=\frac{1}{rr'r''}\int^{\infty}_{0}\frac{\sin{kr''}}{2k}\left[\cos k(r-r')-\cos k(r+r')\right]
\end{align}
where the spherical Bessel function reduces to sinc function for $l=0$
\begin{align}
j_{0}(kr)=\sin(kr)=\frac{\sin{kr}}{kr}
\end{align}
lastly, one can employ the following identity
\begin{align}
\int^{\infty}_{0}\text{d}k''\frac{\sin{kr''}}{k}\cos(kr)=\Theta(k''-k)
\end{align}
where the value of integral is specified to $1/2$ at $r'=r$hence, the averaged density reduces to equation \eqref{Eq.nbar_formula}. 

\subsection{angle-averaged density for truncated potential}\label{Appendix:density_phi12}
\subsubsection{the identity 1}
The present section shows the following mathematical identity
\begin{align}
I_{1}\equiv\int \hat{r'}n(\bmath{r}-\bmath{r'})\text{d}\Omega'=\frac{4\pi}{3}a_{1}(r,r')\hat{r'}\label{Eq.iden_1}
\end{align}
Employing the addition theorem for the Legendre polynomials
\begin{align}
&I_{1}=\sum_{l=0}^{\infty}a_{l}(r,r')\text{e}^{im\phi_{r}}P^{m}_{l}\left(\cos\theta_{r}\right)\frac{(l-m)!}{(l+m)!}\bmath{Q}\\
&\bmath{Q}\equiv\int\text{d}\Omega'\hat{r'}\text{e}^{-im\phi_{r'}}P^{m}_{l}\left(\cos\theta_{r'}\right)
\end{align}
Making use of the basic properties of the associated Legendre polynomials
\begin{align}
&P^{-m}_{l}(x)=(-1)^{m}\frac{(l-m)!}{(l+m)!}P^{m}_{l}(x)\\
&\int^{1}_{-1}\text{d}xP^{k}_{l}(x)P^{m}_{l}(x)=\frac{2(l+k)!}{(2l+1)(l-m)!}\delta_{l,k}
\end{align}
one can obtain the following form
\begin{align}
\bmath{Q}_{l,m}=\frac{4\pi}{3}\delta_{l,1}\left(\frac{\delta_{m,-1}}{2}-\delta_{m,1}, -\frac{\delta_{m,-1}}{2}-\delta_{m,1},\delta_{m,0}\right)
\end{align}
Then one obtains equation \eqref{Eq.iden_1}.

\subsubsection{averaged density}
Employing \eqref{Eq.iden_1}, the angle-averaged density for truncated potential reads
\begin{align}
n_{\Phi}(r)=n_{\bmath{A}}-\frac{\triangle}{3}\nabla\cdot\left[a_{1}(r,r')\hat{r}\right]
\end{align}
where 
\begin{align}
&a_{1}(r,r')=\frac{6}{\pi}\int\text{d}\bmath{r}''r''^{3}n(r'')J_{011}(r,r',r'')\\
&J_{011}(r,r',r'')=\int\text{d}k^{2}j_{1}(kr)j_{1}(kr')j_{0}(kr'')
\end{align}
Since the calculation of higher order for the function $J_{0nn}$ is more tedious, the formula is just given \citep[See e.g.][]{Mehrem_2011}
\begin{align}
J_{011}(r,r',r'')=\frac{\pi(r^{2}+r'^{2}-r''^{2})}{8r^{2}r'^{2}r''}k^{2}\Theta(r+r'-r'')\Theta(r''+r'-r)
\end{align}
Hence, the function is
\begin{align}
a_{1}(r,r')=\frac{3}{4r^{2}r'^{2}}\left(-\int^{r+r'}_{|r-r'|}r''^{3}n(r'')\text{d}r''+[r^{2}+r'^{2}]\int^{r+r'}_{|r-r'|}r''n(r'')\text{d}r''\right)
\end{align}
After some calculation, one obtains 
\begin{align}
n_{\Phi}(r)=&n_{\bmath{A}}-\frac{1}{2r\triangle}\int^{r+r'}_{|r-r'|}r''n(r'')\text{d}r''\nonumber\\
&+\frac{(r+\triangle)n(r+\triangle)+(r-\triangle)n(\mid r-\triangle\mid)}{2r}
\end{align}
Hence, from the definition for the angle-averaged density for truncated acceleraiton, one obtains equation \eqref{Eq.n_Phi}. 
\end{appendices}


\bsp	
\label{lastpage}
\end{document}